\documentclass[fleqn,usenatbib]{mnras}

\usepackage{newtxtext,newtxmath}

\usepackage[T1]{fontenc}
\usepackage{ae,aecompl}

\usepackage{graphicx}	
\usepackage{amsmath}	
\usepackage{amssymb}	

\newcommand{\Ni}{\ensuremath{^{56}\mathrm{Ni}}}

\newcommand{\Msun}{\ensuremath{\mathrm{M}_\odot}}
\newcommand{\Rsun}{\ensuremath{\mathrm{R}_\odot}}
\newcommand{\Msunpyr}{\ensuremath{\mathrm{M}_\odot}~\mathrm{yr^{-1}}}
\newcommand{\vwind}{\ensuremath{v_\mathrm{wind}}}
\newcommand{\Mdot}{\ensuremath{\dot{M}}}
\newcommand{\kmps}{\ensuremath{\mathrm{km~s^{-1}}}}

\newcommand{\rhocsm}{\ensuremath{\rho_\mathrm{CSM}}}

\title[SN IIP LCs with accelerated RSG winds]{
Type~IIP supernova light curves affected by the acceleration of red supergiant winds
}

\author[T. J. Moriya et al.]{
Takashi J. Moriya,$^{1}$\thanks{E-mail: takashi.moriya@nao.ac.jp (TJM)}\thanks{NAOJ Fellow}
Francisco F\"orster,$^{2,3}$
Sung-Chul Yoon,$^{4,5}$ 
G\"otz Gr\"afener,$^{6}$ \newauthor and
Sergei I. Blinnikov$^{7,8,9}$
\\
$^{1}$Division of Theoretical Astronomy, National Astronomical Observatory of Japan, 
National Institutes of Natural Sciences, \\ 2-21-1 Osawa, Mitaka, Tokyo 181-8588, Japan\\
$^{2}$
Center for Mathematical Modelling, University of Chile, Av. Blanco Encalada 2120, Piso 7, Santiago, Chile \\
$^{3}$
Millennium Institute of Astrophysics (MAS), Nuncio Monse\~nor S\'otero Sanz 100, Providencia, Santiago, Chile \\
$^{4}$
Department of Physics and Astronomy, Seoul National University, 08826, Seoul, South Korea \\
$^{5}$
Monash Centre for Astrophysics, School of Physics and Astronomy, Monash University, Victoria 3800, Australia \\
$^{6}$
Argelander Institute for Astronomy, University of Bonn, Auf dem H\"ugel 71, D-53121 Bonn, Germany \\
$^{7}$
Institute for Theoretical and Experimental Physics, Bolshaya Cheremushkinskaya ulitsa 25, 117218 Moscow, Russia \\
$^{8}$
All-Russia Research Institute of Automatics, Sushchevskaya ulitsa 22, 127055 Moscow, Russia \\
$^{9}$
Kavli Institute for the Physics and Mathematics of the Universe (WPI),
The University of Tokyo Institutes for Advanced Study, \\
The University of Tokyo, 5-1-5 Kashiwanoha, Kashiwa, Chiba 277-8583, Japan 
}

\date{Accepted 2018 February 19. Received 2018 February 13; in original form 2017 November 21}

\pubyear{2018}

\begin{document}
\label{firstpage}
\pagerange{\pageref{firstpage}--\pageref{lastpage}}
\maketitle

\begin{abstract}
We introduce the first synthetic light-curve model set of Type~IIP supernovae exploded within circumstellar media in which the acceleration of the red supergiant winds is taken into account. Because wind acceleration makes the wind velocities near the progenitors low, the density of the immediate vicinity of the red supergiant supernova progenitors can be higher than that extrapolated by using a constant terminal wind velocity. Therefore, even if the mass-loss rate of the progenitor is relatively low, it can have a dense circumstellar medium at the immediate stellar vicinity and the early light curves of Type~IIP supernovae are significantly affected by it. We adopt a simple $\beta$ velocity law to formulate the wind acceleration. We provide bolometric and multicolor light curves of Type~IIP supernovae exploding within such accelerated winds from the combinations of three progenitors, $12-16$~\Msun; five $\beta$, $1-5$; seven mass-loss rates, $10^{-5}-10^{-2}~\Msunpyr$; and four explosion energies, $(0.5-2)\times 10^{51}~\mathrm{erg}$. All the light curve models are available at \url{https://goo.gl/o5phYb}. When the circumstellar density is sufficiently high, our models do not show a classical shock breakout as a consequence of the interaction with the dense and optically-thick circumstellar media. Instead, they show a delayed 'wind breakout', substantially affecting early light curves of Type~IIP supernovae. We find that the mass-loss rates of the progenitors need to be $10^{-3}-10^{-2}~\Msunpyr$ to explain typical rise times of $5-10$~days in Type~IIP supernovae assuming a dense circumstellar radius of $10^{15}~\mathrm{cm}$.
\end{abstract}




\section{Introduction}
Type IIP supernovae (SNe) are explosions of red supergiants (RSGs) (see \citealt{smartt2015review} for a review). RSGs are known to have the mass-loss rates of up to around $10^{-4}~\Msunpyr$ \citep[e.g.,][]{goldman2017agbrsgwind}, which may occasionally increase to $\sim 10^{-3}~\Msunpyr$ \citep[e.g.,][]{smith2009vycma}. Radio observations of SNe~IIP more than $\sim 10~\mathrm{days}$ after the explosions has revealed that the mass-loss rates of SN~IIP progenitors are similar to those of regular RSGs \citep[e.g.,][]{chevalier2006typeiipcsm}.

However, recent early observations of SNe~IIP are revealing the existence of unexpectedly dense circumstellar environments very close to SN~IIP progenitors. The possible existence of high-density circumstellar media (CSM) at the immediate vicinity of the RSG SN progenitors has been suggested based on the early SN~IIP LC observations. For example, the rise times of SNe~IIP are often much shorter than those predicted from theoretical LC modeling \citep[e.g.,][]{gonzalez2015iiprise,gall2015earlyiip,garnavich2016keplerbreakout,rubin2016iiprise,rubin2017kepler}. One possible suggested reason for SNe~IIP to have a quick rise is the existence of dense CSM \citep[e.g.,][]{falk1977arnett2,moriya2011iipcsm}. SN ejecta clash into the dense CSM and the collision results in the efficient conversion of the kinetic energy to radiation, making SNe to rise fast. There are several SN~IIP LC signatures that are suggested to be caused by the dense CSM \citep[e.g.,][]{nagy2016iipbump,morozova2017iil,morozova2017iipint2}.

The direct evidence for the existence of the dense CSM around SN~IIP progenitors is found in their early time spectra. \citet{khazov2016earlyiispectra} found that narrow emission lines that indicate the existence of the dense CSM are often seen in SNe~IIP when they are observed early enough. The earliest spectrum of SNe~IIP taken so far is at around 6~hours after the explosion of SN~2013fs \citep{yaron2017iipcsm}. It showed strong narrow hydrogen emission lines with broad electron scattering wings that indicate the existence of the dense CSM around the progenitor. By modeling the early spectra, \citet{yaron2017iipcsm} conclude that the progenitor had a mass-loss rate of $\sim 10^{-4}-10^{-3}~\Msunpyr$ where we assume a constant wind velocity of $10~\kmps$. However, the strong emission lines disappeared in about 5~days and the spectra turned into those of usual SNe~IIP. Later radio observations of SN~2013fs show that the CSM that is far from the progenitor $(>5\times 10^{15}~\mathrm{cm})$ is made with the mass-loss rate of $\lesssim 10^{-6}~\Msunpyr$ where we assume a wind velocity of 10~\kmps. \citet{yaron2017iipcsm} conclude that only the immediate vicinity ($\lesssim 10^{15}~\mathrm{cm}$) of the SN~2013fs progenitor is surrounded by the dense CSM and this is likely a result of the increasing mass-loss rate of the progenitor in the final years before the explosion. 

Meanwhile, \citet{morozova2017iil} conducted a LC modeling of SN~2013fs and concluded that the mass-loss rate of the progenitor of SN~2013fs needs to be $\sim 0.1~\Msunpyr$ in the final years to the explosion to explain its LC behavior, assuming a constant wind velocity of 10~\kmps. The estimated rate is far beyond that estimated from the early spectrum.

To reconcile the large inconsistency, \citet{moriya2017windacc13fs} pointed out the possible importance of the wind acceleration in the density structure of the immediate vicinity of SN~IIP progenitors. LC modeling by \citet{morozova2017iil} assume that the mass-loss rate and wind velocity are kept constant at the stellar surface and the CSM density structure follows $r^{-2}$ up to the stellar surface. However, the wind is actually accelerated at the stellar surface gradually and does not have a constant velocity. Because smaller wind velocities make CSM densities higher, the CSM density at the immediate stellar vicinity can be rather high due to the wind acceleration even if the mass-loss rate is relatively low. Indeed, \citet{moriya2017windacc13fs} show that the mass-loss rate of the SN~2013fs progenitor could be as low as $10^{-3}~\Msunpyr$ to explain the early LC properties if the wind acceleration is taken into account. This mass-loss rate is consistent with that estimated from the spectral modeling \citep{yaron2017iipcsm}. The necessity to take the wind acceleration into account is also found in the spectral modeling of the early spectrum of SN~2013cu \citep{gal-yam2014flash}, although SN~2013cu is a Type~IIb SN (\citealt{grafener2016flashmodel}, see also \citealt{groh2014flash}).

In our previous study of the effect of the wind acceleration on early SN~IIP LCs, we focused on the particular case of SN~2013fs. However, as mentioned before, there are a lot of observational indications that the early SN~IIP LCs are commonly affected by dense CSM. In this study, we continue our study on the effect of the wind acceleration on early SN~IIP LCs by performing numerical LC modeling in a wide parameter range of possible wind acceleration as well as SN properties to investigate their general properties. 

The rest of this paper is organized as follows. We present our methods of the LC calculations in Section~\ref{sec:mothods}. We present our LCs in Section~\ref{sec:results}. We discuss and conclude our study in Section~\ref{sec:discussion}. All the LCs presented in this work are available at \url{https://goo.gl/o5phYb}. All the magnitudes in this paper are AB magnitudes.

\begin{table}
	\centering
	\caption{RSG progenitor properties.}
	\label{tab:progenitor}
	\begin{tabular}{cccc} 
		\hline
        & & Final H-rich & \\
		ZAMS mass & Final mass & envelope mass & Final radius\\
		\hline
        12~\Msun & 10.3~\Msun & 6.1~\Msun & 607~\Rsun \\
        14~\Msun & 11.4~\Msun & 6.2~\Msun & 832~\Rsun \\
        16~\Msun & 12.0~\Msun & 5.8~\Msun & 962~\Rsun \\
		\hline
	\end{tabular}
\end{table}

\begin{table}
	\centering
	\caption{CSM mass. The CSM radius is set as $10^{15}~\mathrm{cm}$.}
	\label{tab:csmmass}
	\begin{tabular}{ccccc} 
		\hline
		\Mdot ($\Msunpyr$) &  $\beta$ & \multicolumn{3}{c}{CSM mass (\Msun)}\\
		                   &          & 12~\Msun & 14~\Msun & 16~\Msun \\        
		\hline
       $10^{-5}$ & 0$^a$& 0.0003& 0.0003& 0.0003\\
       $10^{-5}$ & 1    & 0.002 & 0.003 & 0.004 \\
       $10^{-5}$ & 1.75 & 0.004 & 0.006 & 0.008 \\
       $10^{-5}$ & 2.5  & 0.009 & 0.013 & 0.031 \\
       $10^{-5}$ & 3.75 & 0.068 & 0.093 & 0.153 \\
       $10^{-5}$ & 5    & 0.154 & 0.210 & 0.360 \\
$3\times10^{-5}$ & 0$^a$& 0.0009& 0.0009& 0.0009\\
$3\times10^{-5}$ & 1    & 0.003 & 0.004 & 0.006 \\
$3\times10^{-5}$ & 1.75 & 0.007 & 0.010 & 0.012 \\
$3\times10^{-5}$ & 2.5  & 0.027 & 0.038 & 0.046 \\
$3\times10^{-5}$ & 3.75 & 0.097 & 0.132 & 0.217 \\
$3\times10^{-5}$ & 5    & 0.201 & 0.275 & 0.468 \\ 
       $10^{-4}$ & 0$^a$& 0.003 & 0.003 & 0.003 \\
       $10^{-4}$ & 1    & 0.007 & 0.008 & 0.010 \\
       $10^{-4}$ & 1.75 & 0.015 & 0.020 & 0.024 \\
       $10^{-4}$ & 2.5  & 0.089 & 0.129 & 0.153 \\
       $10^{-4}$ & 3.75 & 0.144 & 0.195 & 0.318 \\
       $10^{-4}$ & 5    & 0.276 & 0.375 & 0.633 \\         
$3\times10^{-4}$ & 0$^a$& 0.009 & 0.009 & 0.009 \\
$3\times10^{-4}$ & 1    & 0.015 & 0.018 & 0.020 \\
$3\times10^{-4}$ & 1.75 & 0.030 & 0.048 & 0.054 \\
$3\times10^{-4}$ & 2.5  & 0.080 & 0.106 & 0.198 \\
$3\times10^{-4}$ & 3.75 & 0.213 & 0.287 & 0.460 \\
$3\times10^{-4}$ & 5    & 0.377 & 0.512 & 0.850 \\  
       $10^{-3}$ & 0$^a$& 0.030 & 0.030 & 0.030 \\
       $10^{-3}$ & 1    & 0.044 & 0.049 & 0.053 \\
       $10^{-3}$ & 1.75 & 0.098 & 0.124 & 0.140 \\
       $10^{-3}$ & 2.5  & 0.162 & 0.208 & 0.235 \\
       $10^{-3}$ & 3.75 & 0.346 & 0.459 & 0.714 \\
       $10^{-3}$ & 5    & 0.561 & 0.752 & 1.216 \\
$3\times10^{-3}$ & 0$^a$& 0.091 & 0.090 & 0.089 \\
$3\times10^{-3}$ & 1    & 0.126 & 0.137 & 0.144 \\
$3\times10^{-3}$ & 1.75 & 0.212 & 0.285 & 0.315 \\
$3\times10^{-3}$ & 2.5  & 0.325 & 0.472 & 0.529 \\
$3\times10^{-3}$ & 3.75 & 0.581 & 0.932 & 1.060 \\
$3\times10^{-3}$ & 5    & 0.864 & 1.452 & 1.659 \\    
       $10^{-2}$ & 0$^a$& 0.304 & 0.299 & 0.296 \\
       $10^{-2}$ & 1    & 0.406 & 0.434 & 0.459 \\
       $10^{-2}$ & 1.75 & 0.585 & 0.660 & 0.774 \\
       $10^{-2}$ & 2.5  & 0.783 & 0.974 & 1.252 \\
       $10^{-2}$ & 3.75 & 1.193 & 1.571 & 2.191 \\
       $10^{-2}$ & 5    & 1.626 & 2.204 & 3.209 \\    
        \hline
\multicolumn{5}{l}{$^a$$\beta=0$ is the model without the wind acceleration}\\
\multicolumn{5}{l}{where $\rhocsm\propto r^{-2}$ is achieved at the stellar surface.}\\ 
	\end{tabular}
\end{table}

\section{Methods}\label{sec:mothods}
\subsection{Progenitors}\label{sec:progenitors}
The RSG SN progenitors are obtained by using the public stellar evolution code \texttt{MESA} \citep{paxton2011mesa1,paxton2013mesa2,paxton2015mesa3,paxton2017mesa4} as in \citet{moriya2017windacc13fs}. Briefly, we adopt the Ledoux criterion for convection with a mixing-length parameter of 2.0 and a semiconvection parameter of 0.01. Overshooting is taken into account on top of the hydrogen-burning convective core with a step function. We use the overshooting parameter of $0.3 H_P$, where $H_P$ is the pressure scale height. The standard `Dutch' wind is adopted without scaling for both hot and cool stars. All the models in this study have the solar metallicity at the beginning. They are evolved to the core oxygen burning stage, from which the hydrogen-rich envelope structure hardly changes until the core collapse. We adopt three zero-age main-sequence (ZAMS) masses for this study; 12, 14, and 16~\Msun. The choice of the ZAMS masses is based on the study of \citet{smartt2009rsgprogmass} which shows that the maximum mass of SN~IIP progenitors is at around 16.5~\Msun. The final progenitor properties are summarized in Table~\ref{tab:progenitor}.

The CSM structure is attached on top of the aforementioned RSG progenitors. The CSM density \rhocsm\ is expressed as $\rhocsm(r) = \Mdot/(4\pi\vwind)r^{-2}$, where \Mdot\ is the progenitor's mass-loss rate and \vwind\ is the wind velocity. If both \Mdot\ and \vwind\ are constant, \rhocsm\ is simply proportional to $r^{-2}$. In this study, \Mdot\ is set constant in all the progenitor models but we take the radial change of \vwind\ caused by the wind acceleration into account. The wind does not achieve the final velocity at the stellar surface instantaneously and it gradually acquires its terminal velocity. The wind acceleration mechanisms of RSGs are unknown and there are no established prescriptions to express the velocity distribution around RSGs. Therefore, we adopt a simple $\beta$ velocity law for the wind velocity to account for the wind acceleration in this study, i.e.,
\begin{equation}
v_\mathrm{wind} (r) = v_0 + (v_\infty - v_0) \left( 1 - \frac{R_0}{r} \right)^\beta,
\end{equation}
where $v_0$ is the initial wind velocity, $v_\infty$ is the terminal wind velocity, and $R_0$ is the wind launching radius that we set at the stellar surface. For simplicity, we fix $v_\infty = 10~\mathrm{km~s^{-1}}$. $v_0$ is chosen so that the CSM density is smoothly connected from the surface of the progenitors. The exact value of $v_0$ depends on the progenitor, \Mdot, and $\beta$ but they are always less than $10^{-2}~\mathrm{km~s^{-1}}$ in our models. We take $\beta$ between 1 and 5. OB stars have $\beta\simeq 0.5-1$ \citep[e.g.,][]{groenewegen1989ostarmassloss,haser1995ostarmassloss,puls1996ostarmassloss} and RSGs are known to experience slower wind acceleration than OB stars, i.e., $\beta \gtrsim 1$ \citep[e.g.,][]{bennett2010rsgwind,marshall2004agbwindacc}. For example, a RSG $\zeta$ Aurigae is measured to have $\beta\simeq 3.5$ \citep{schroeder1985windbeta,baade1996wind}. \citet{moriya2017windacc13fs} find that the progenitor of SN~IIP 2013fs may have had $\beta\simeq 5$ based on its LC. We adopt \Mdot\ between $10^{-5}$ and $10^{-2}~\Msun~\mathrm{yr^{-1}}$. In our standard models, the CSM radius is fixed to $10^{15}~\mathrm{cm}$ but we also investigate LC models with the CSM radii of $10^{14}~\mathrm{cm}$, $3\times10^{14}~\mathrm{cm}$, and $3\times 10^{15}~\mathrm{cm}$ (Section~\ref{sec:radii}).

\begin{figure*}
   \includegraphics[width=0.9\columnwidth]{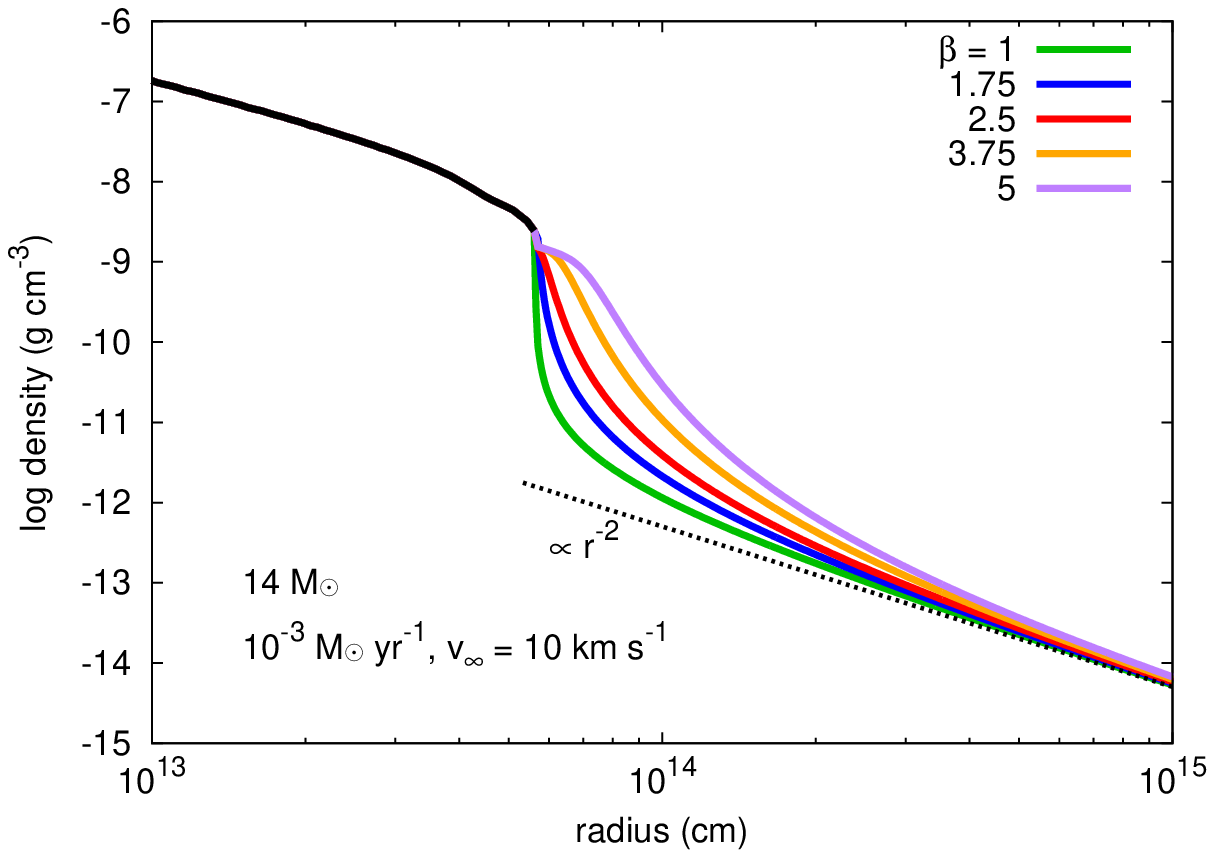}
   \includegraphics[width=0.9\columnwidth]{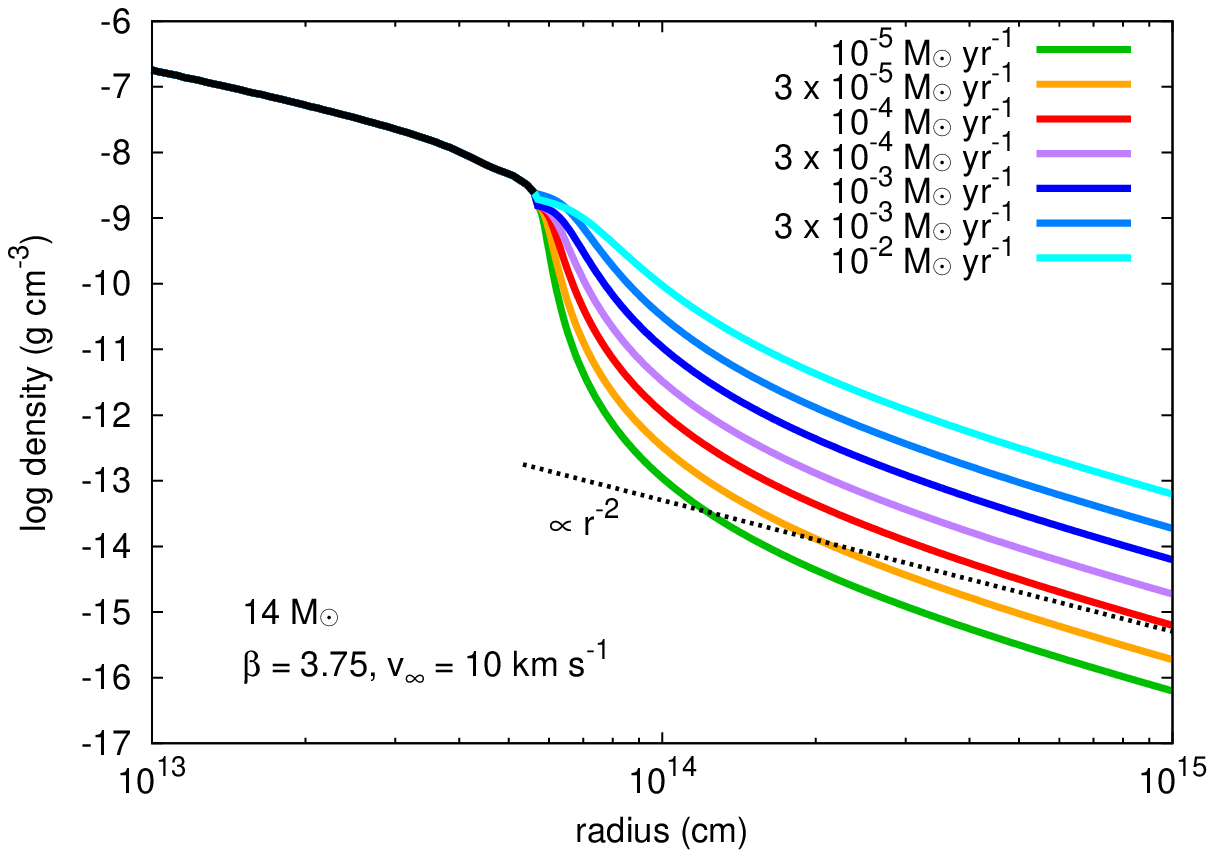}   
   \caption{
   Density structure of the 14~\Msun\ progenitor with CSM that takes the wind acceleration into account.
   \textit{Left:} Density structure with the constant mass-loss rate of $10^{-3}~\Msunpyr$ with different $\beta$. The dotted line is the CSM structure which assumes that the wind is instantaneously accelerated to the terminal velocity $(v_\infty=10~\kmps)$ at the stellar surface and it follows $\propto r^{-2}$. \textit{Right:} Density structure with the constant $\beta=3.75$ with different \Mdot. The dotted line is the $\Mdot=10^{-4}~\Msunpyr$ CSM model that assumes the instantaneous wind acceleration to the terminal velocity of 10~\kmps\ and follows $\rhocsm \propto r^{-2}$.
   }
    \label{fig:density}
\end{figure*}

Fig.~\ref{fig:density} shows some examples of the density structures with different $\beta$ and different \Mdot. The dotted structure ($\rhocsm\propto r^{-2}$) does not take the wind acceleration into account. We can see that the $\beta$ velocity law makes the CSM density at the immediate vicinity of the SN progenitors much larger than that without the acceleration. Table~\ref{tab:csmmass} summarizes the CSM masses in the CSM within $10^{15}~\mathrm{cm}$. For example, the models with $\Mdot=3\times 10^{-4}~\Msunpyr$ that do not take the wind acceleration into account have a CSM mass of 0.009~\Msun. The same progenitor can have a similar amount of CSM with $\Mdot=10^{-5}~\Msunpyr$ if $\beta\gtrsim 2.5$ (Table~\ref{tab:csmmass}). $\beta\gtrsim 2.5$ is consistent with those found in RSGs and RSG SN progenitors are natural to have higher CSM density near the stellar surface than that simply presumed from their mass-loss rates. The CSM mass is an important parameter determining the LC properties of SNe powered by the interaction and the resulting LCs can be similar in the two CSM even though the mass-loss rates are quite different.

\begin{figure*}
   \includegraphics[width=0.9\columnwidth]{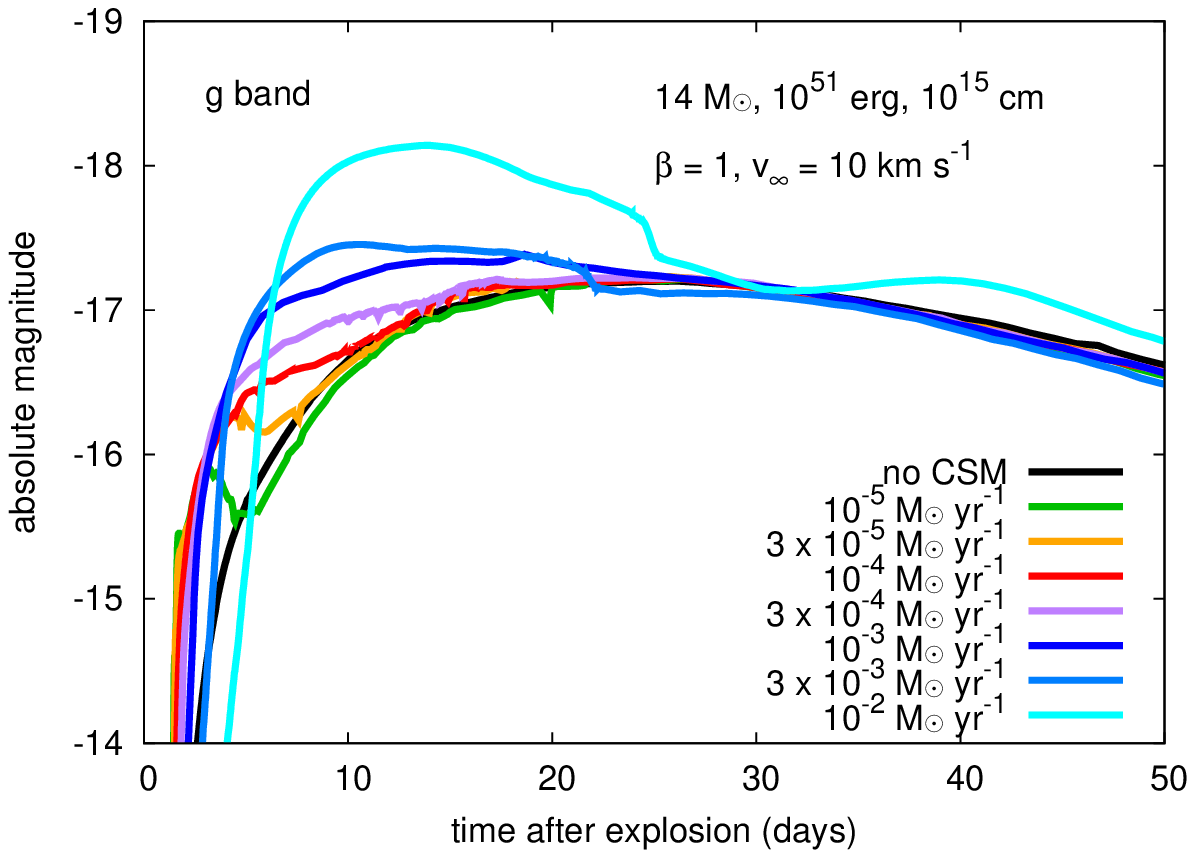}
   \includegraphics[width=0.9\columnwidth]{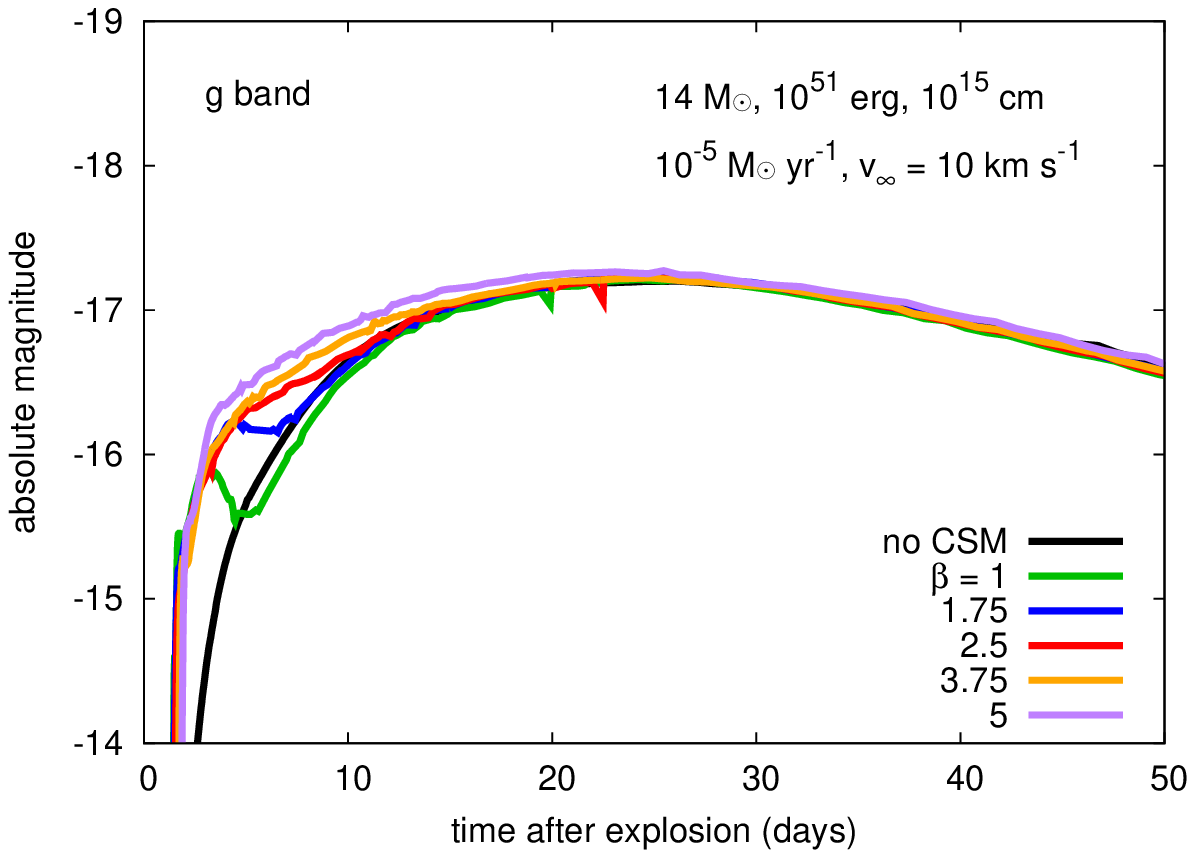} \\
   \includegraphics[width=0.9\columnwidth]{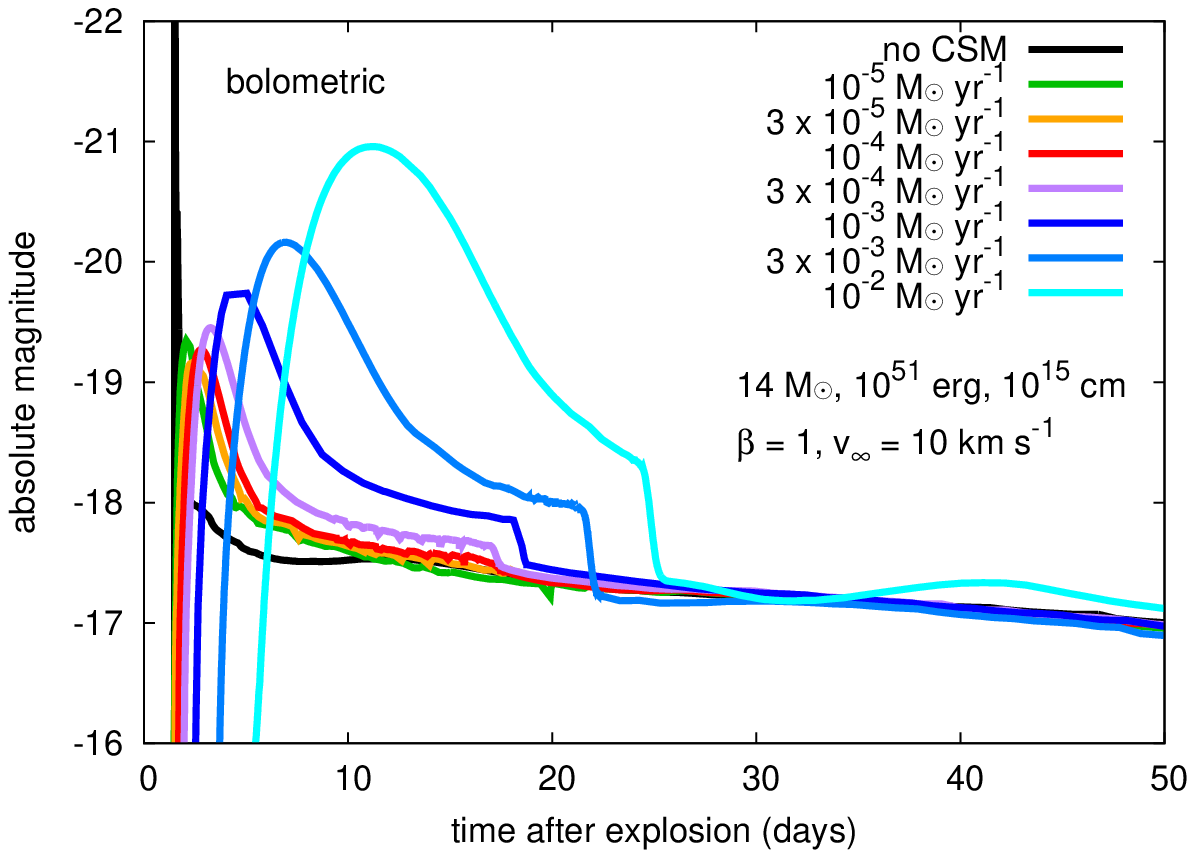}
   \includegraphics[width=0.9\columnwidth]{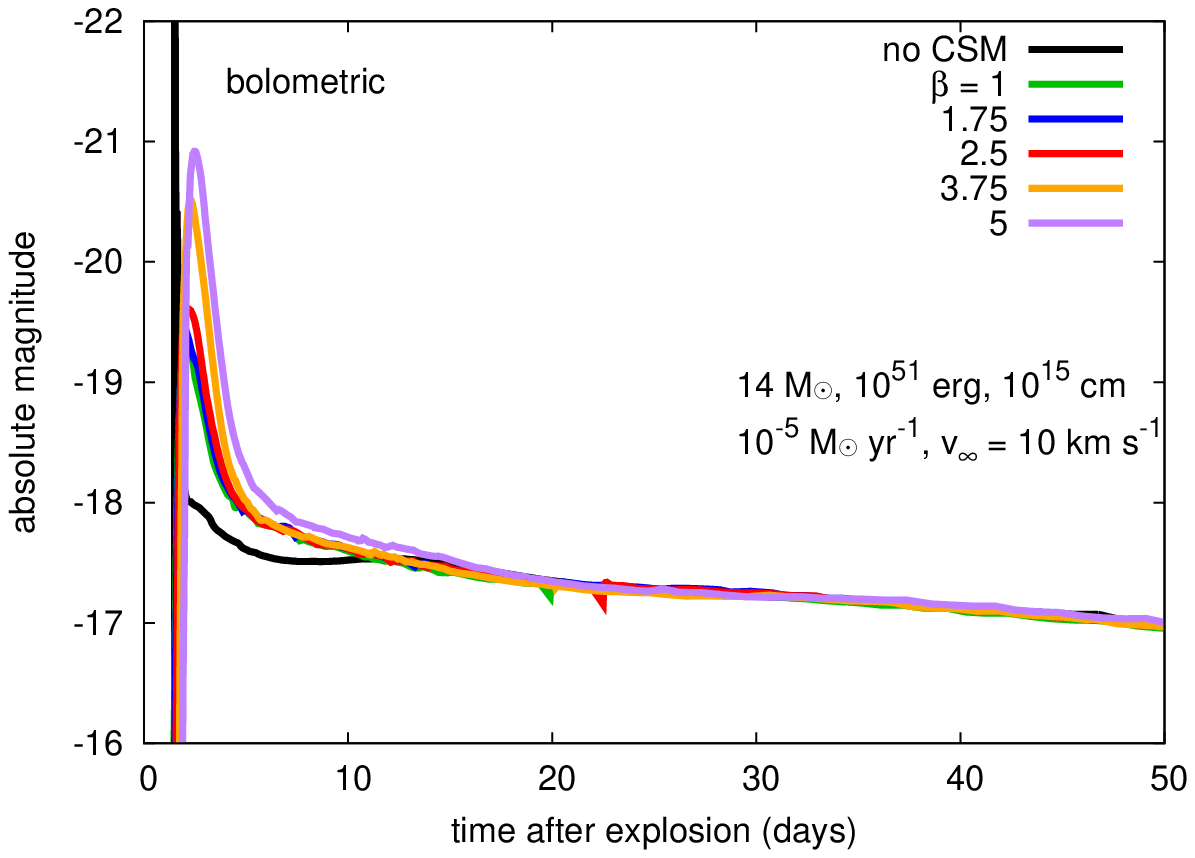}   
   \caption{
   Representative LCs with different \Mdot\ and $\beta$ from the 14~\Msun\ progenitor exploded with the energy of $10^{51}~\mathrm{erg}$. The CSM radius is set as $10^{15}~\mathrm{cm}$. See Table~\ref{tab:csmmass} for the CSM masses of the models. The top panels show the $g$-band LCs and the bottom panel show the bolometric LCs. The left panels show the LCs with a fixed $\beta=1$ and different mass-loss rates, while the right panels show LCs with a fixed $\Mdot=10^{-5}~\Msunpyr$ and different $\beta$.
   }
    \label{fig:lc_lows}
\end{figure*}

\begin{figure*}
   \includegraphics[width=0.9\columnwidth]{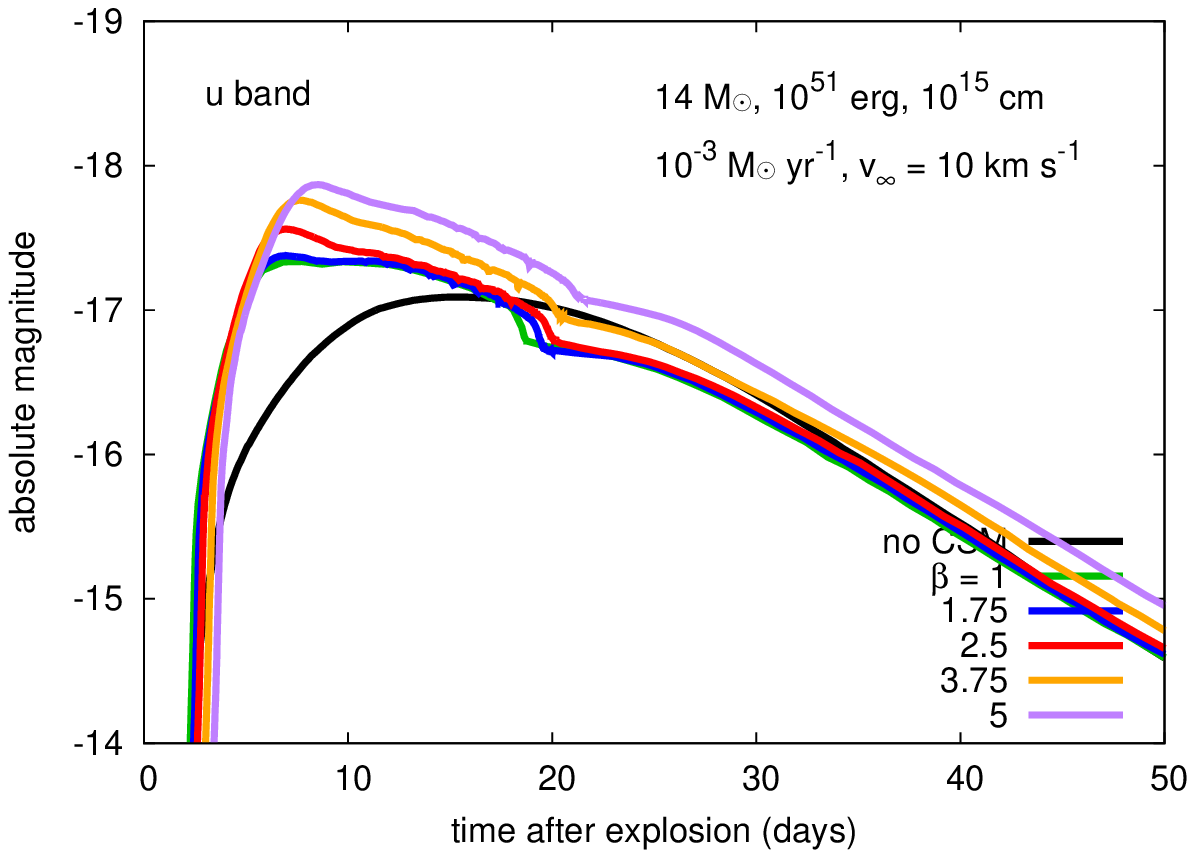}
   \includegraphics[width=0.9\columnwidth]{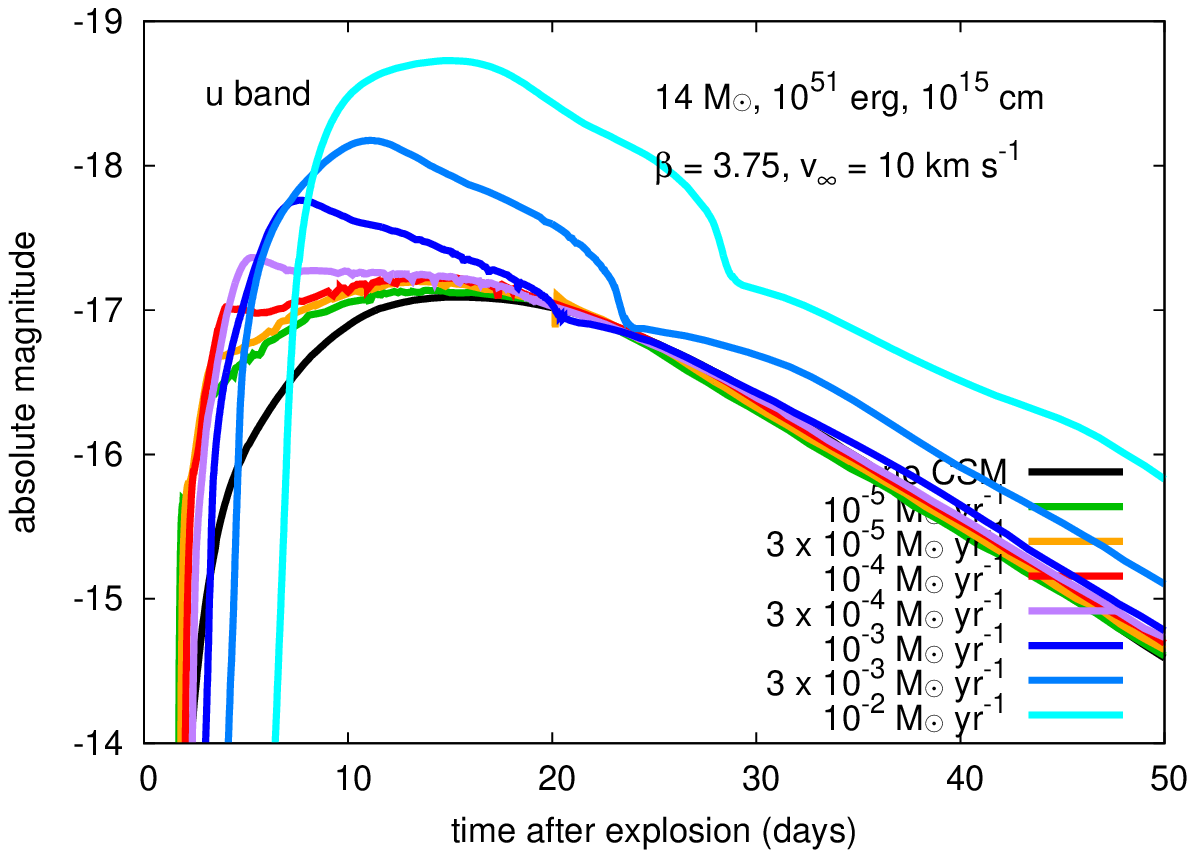} \\
   \includegraphics[width=0.9\columnwidth]{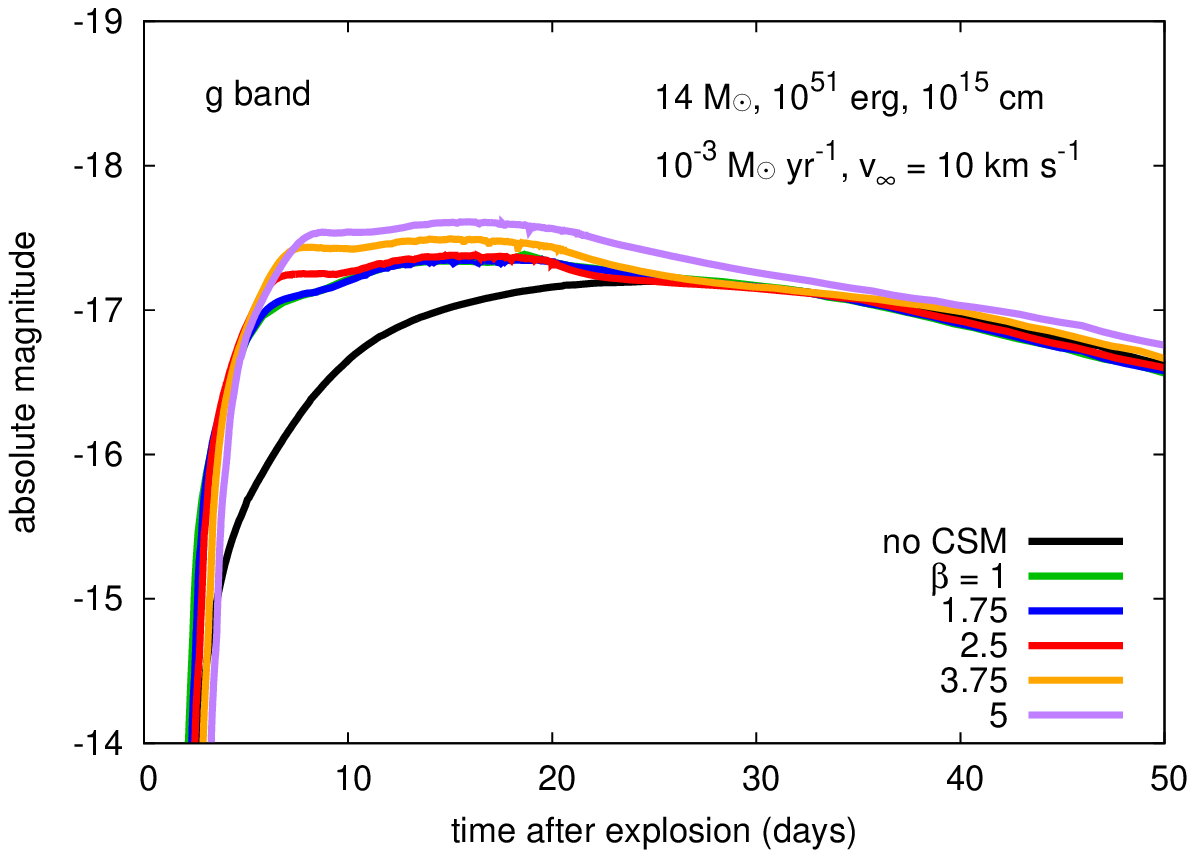}
   \includegraphics[width=0.9\columnwidth]{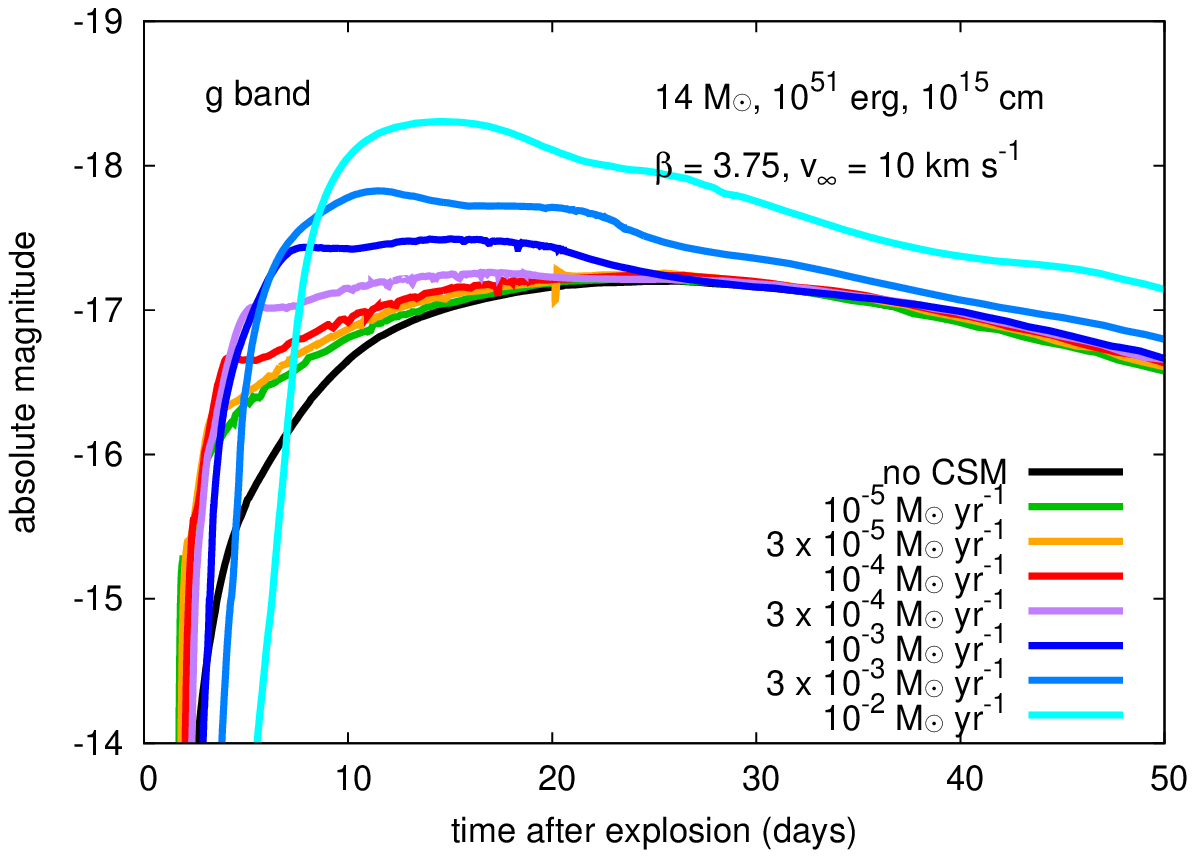} \\
   \includegraphics[width=0.9\columnwidth]{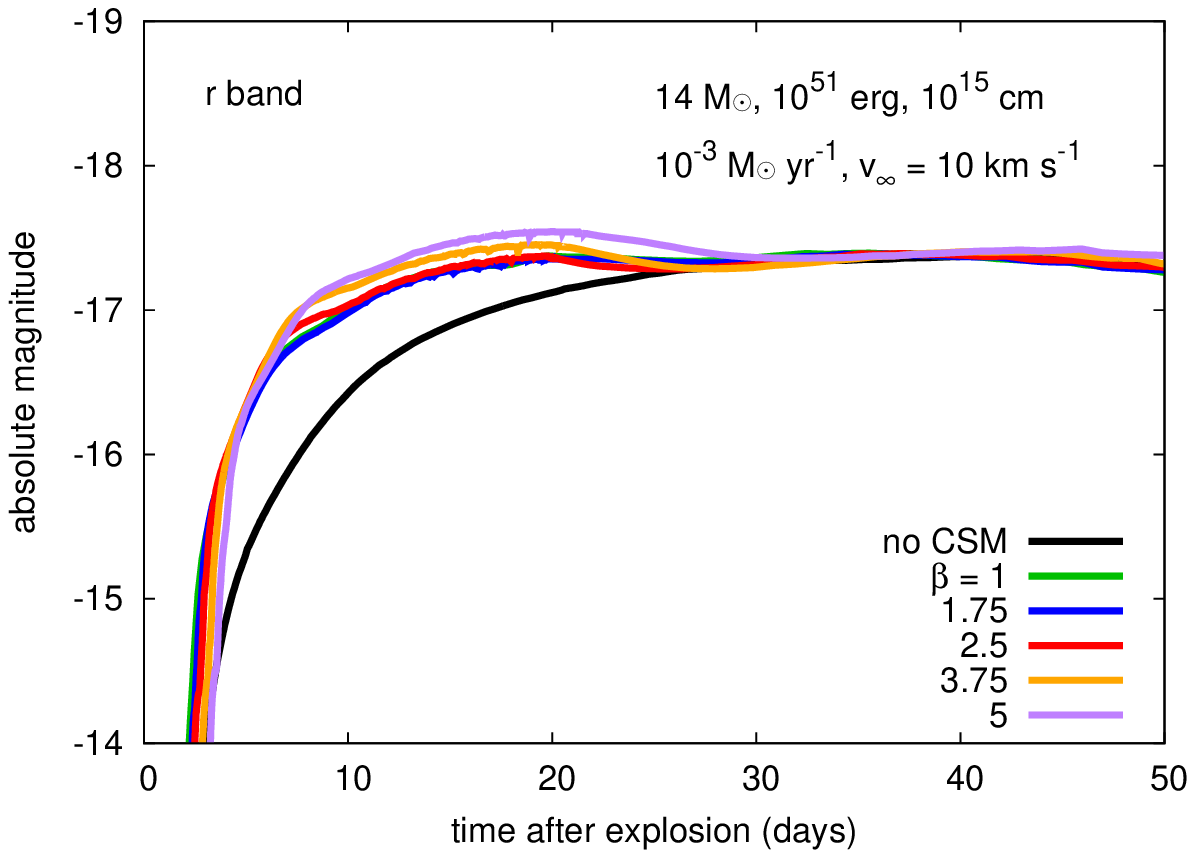}
   \includegraphics[width=0.9\columnwidth]{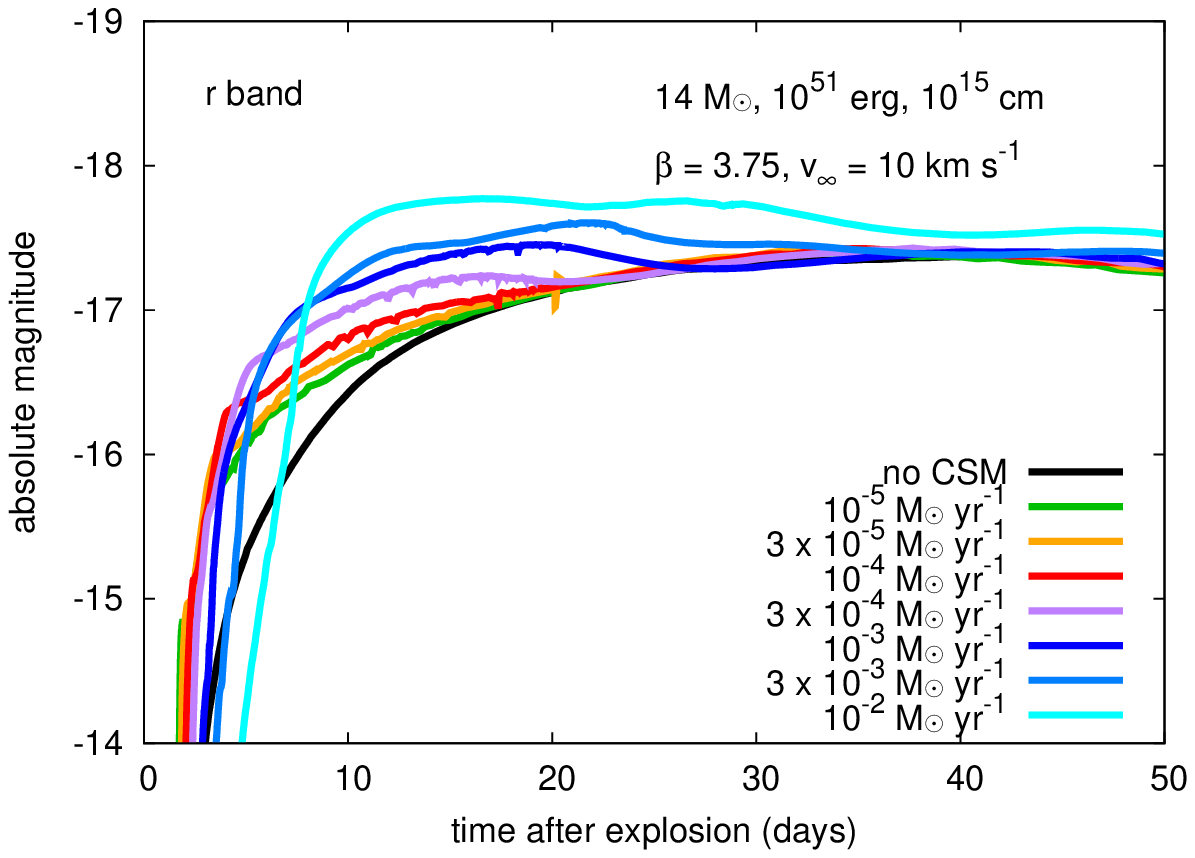} \\
   \includegraphics[width=0.9\columnwidth]{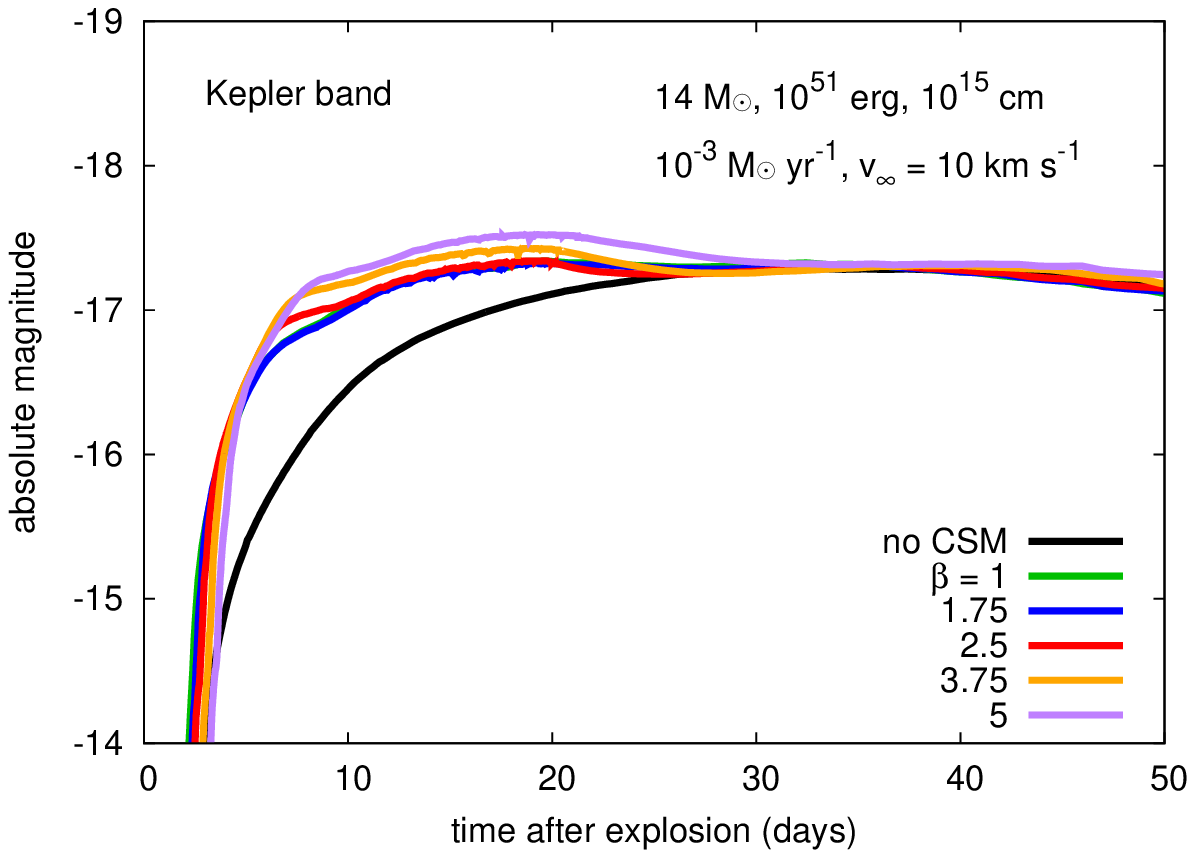}
   \includegraphics[width=0.9\columnwidth]{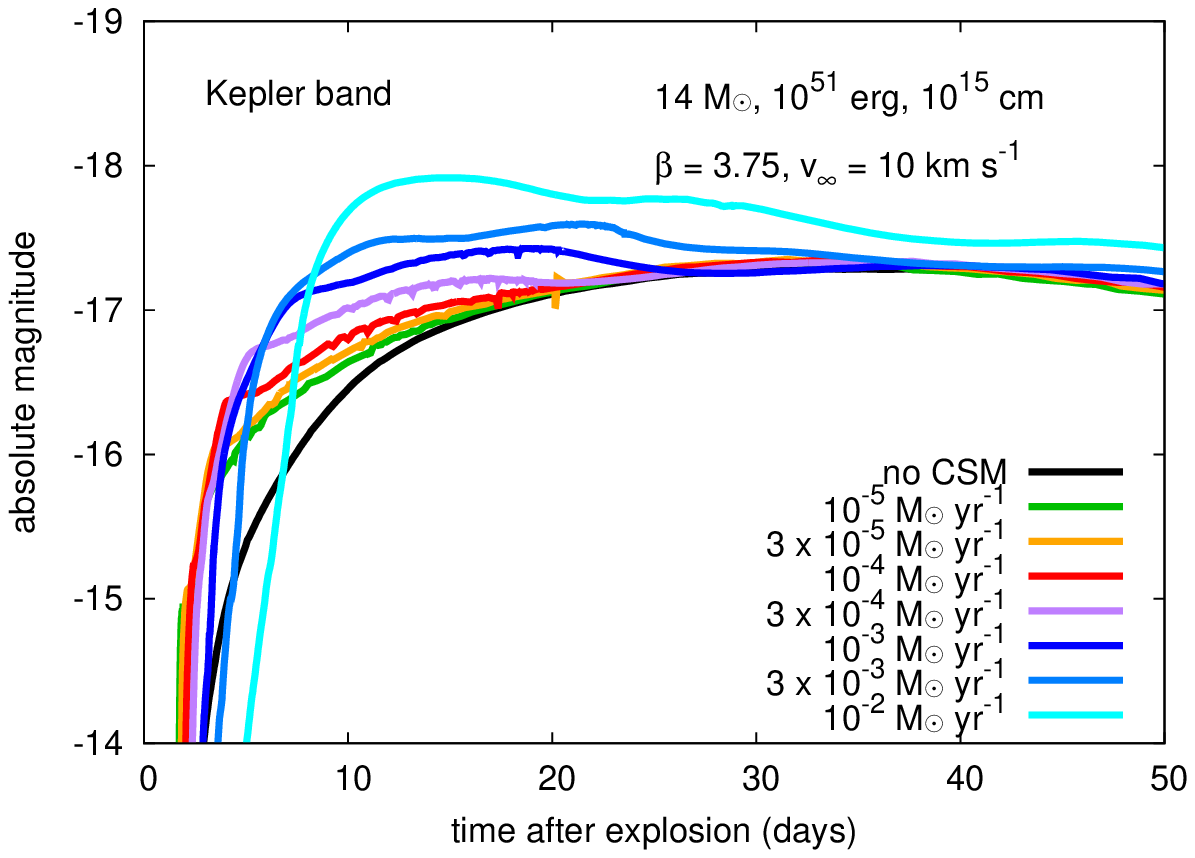}
   \caption{
   Synthetic multicolor LCs. The left panels show the models with the CSM of $\Mdot=10^{-3}~\Msunpyr$ with different $\beta$. The right panels show the models with the CSM of $\beta=3.75$ with different \Mdot. The progenitor model (14~\Msun), the explosion energy ($10^{51}~\mathrm{erg}$), and the dense CSM radius ($10^{15}~\mathrm{cm}$) are fixed.
   }
   \label{fig:lc_multiclc}
\end{figure*}

\begin{figure}
   \includegraphics[width=0.9\columnwidth]{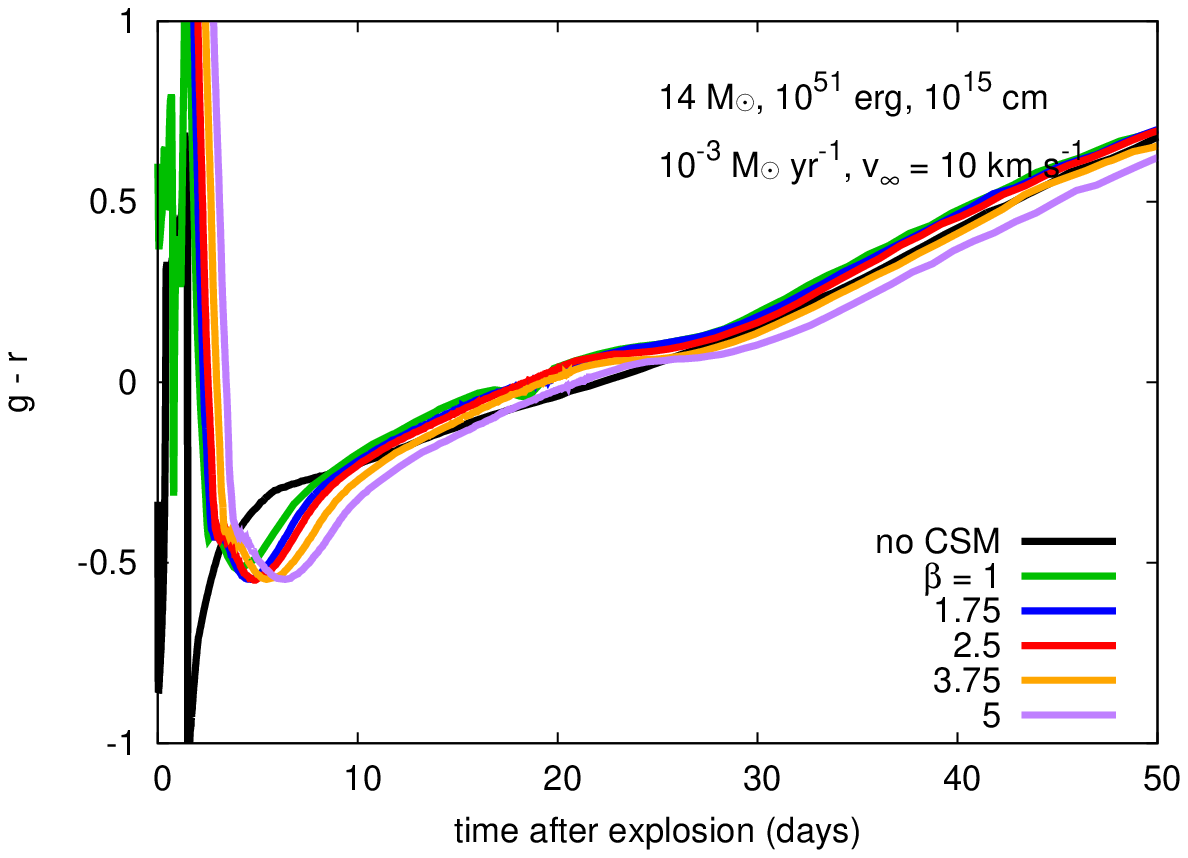}
   \includegraphics[width=0.9\columnwidth]{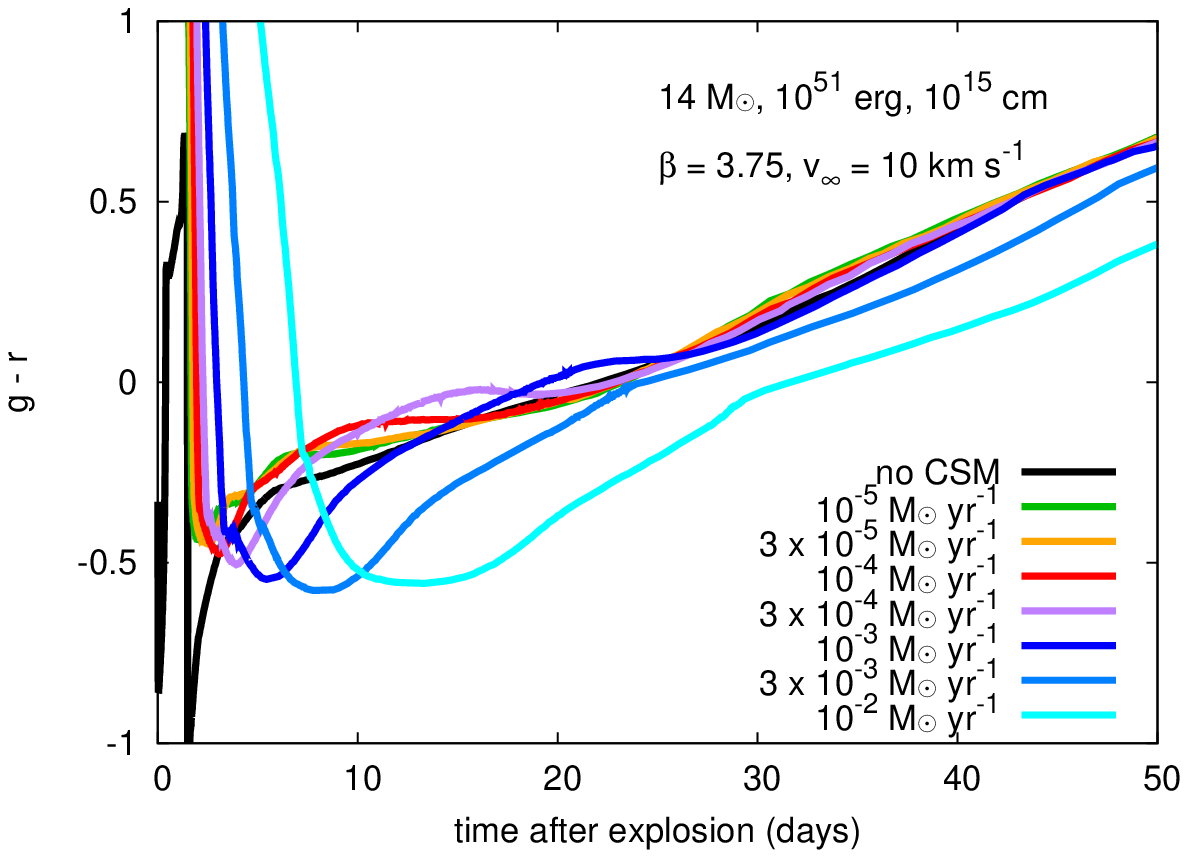}
   \caption{
   Color ($g-r$) evolution of the representative models presented in Fig.~\ref{fig:lc_multiclc}.
   }
    \label{fig:color}
\end{figure}

\begin{figure*}
   \includegraphics[width=0.9\columnwidth]{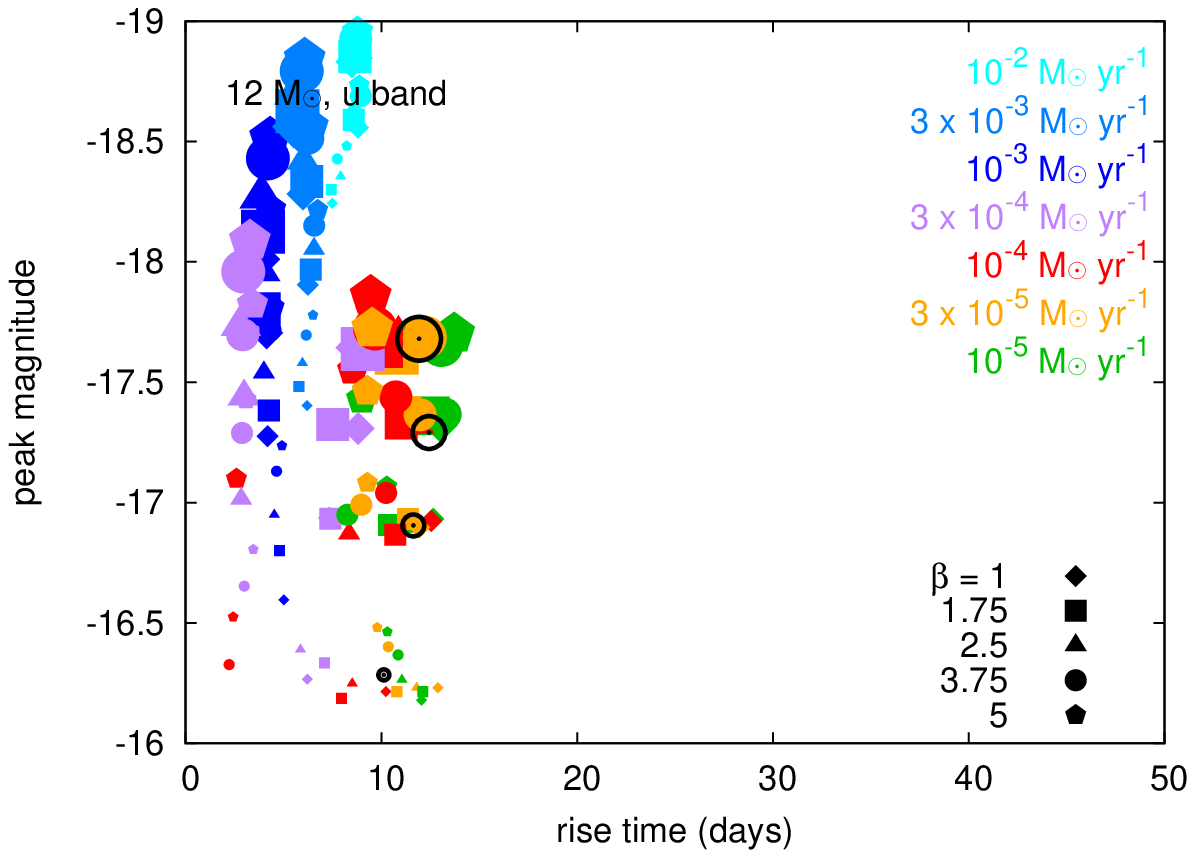}
   \includegraphics[width=0.9\columnwidth]{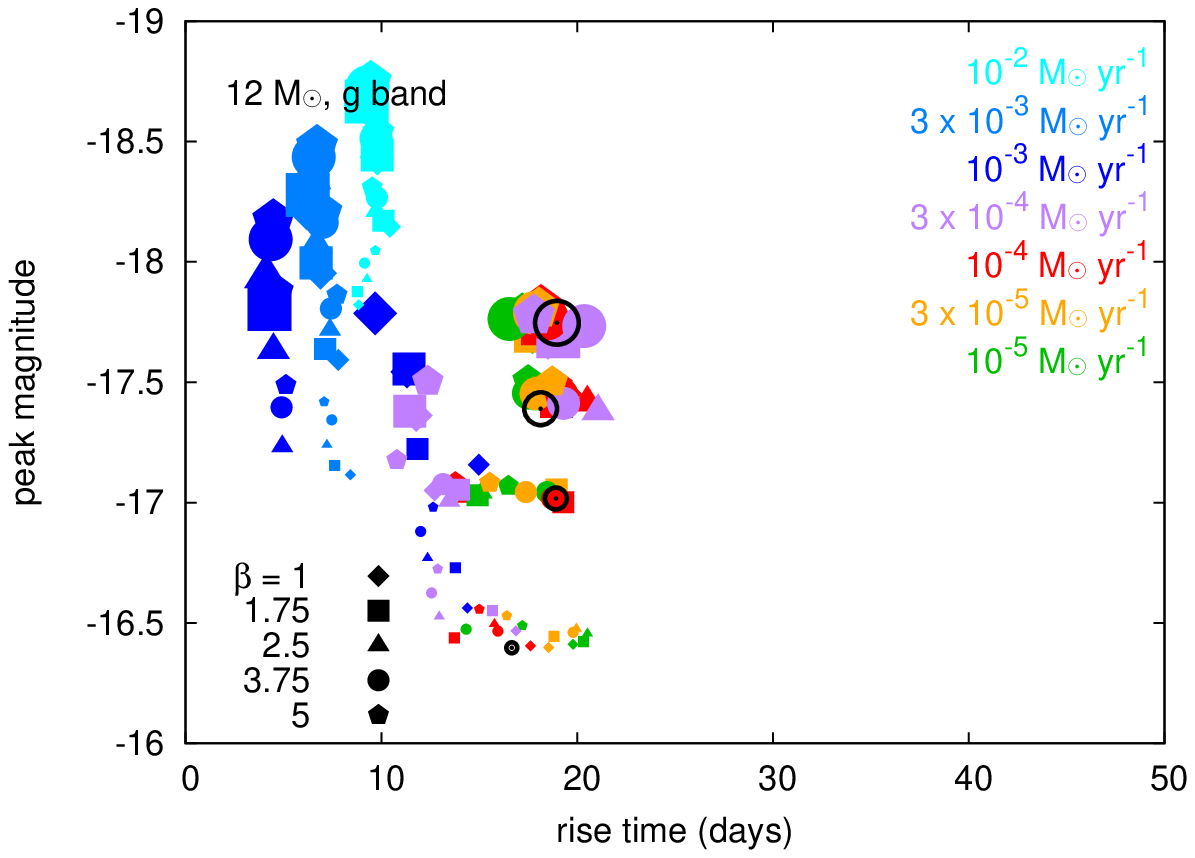} \\
   \includegraphics[width=0.9\columnwidth]{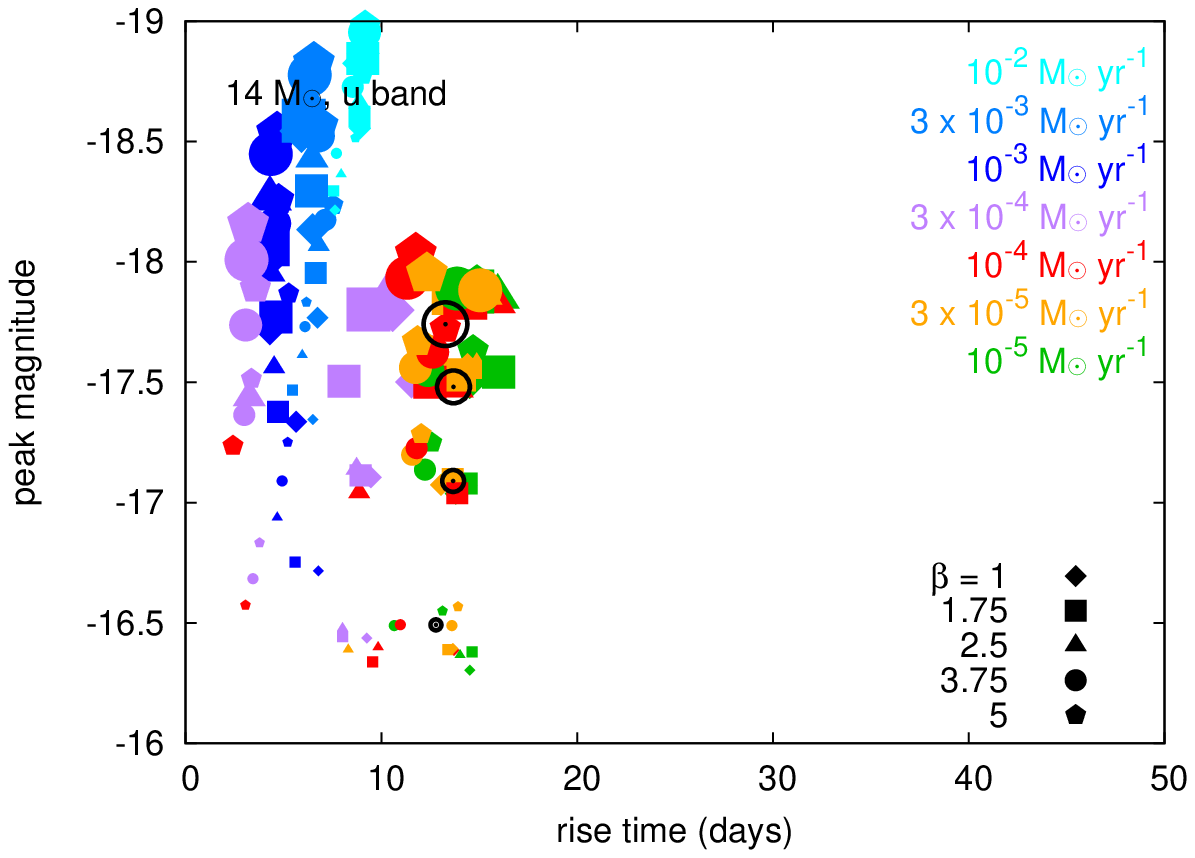}
   \includegraphics[width=0.9\columnwidth]{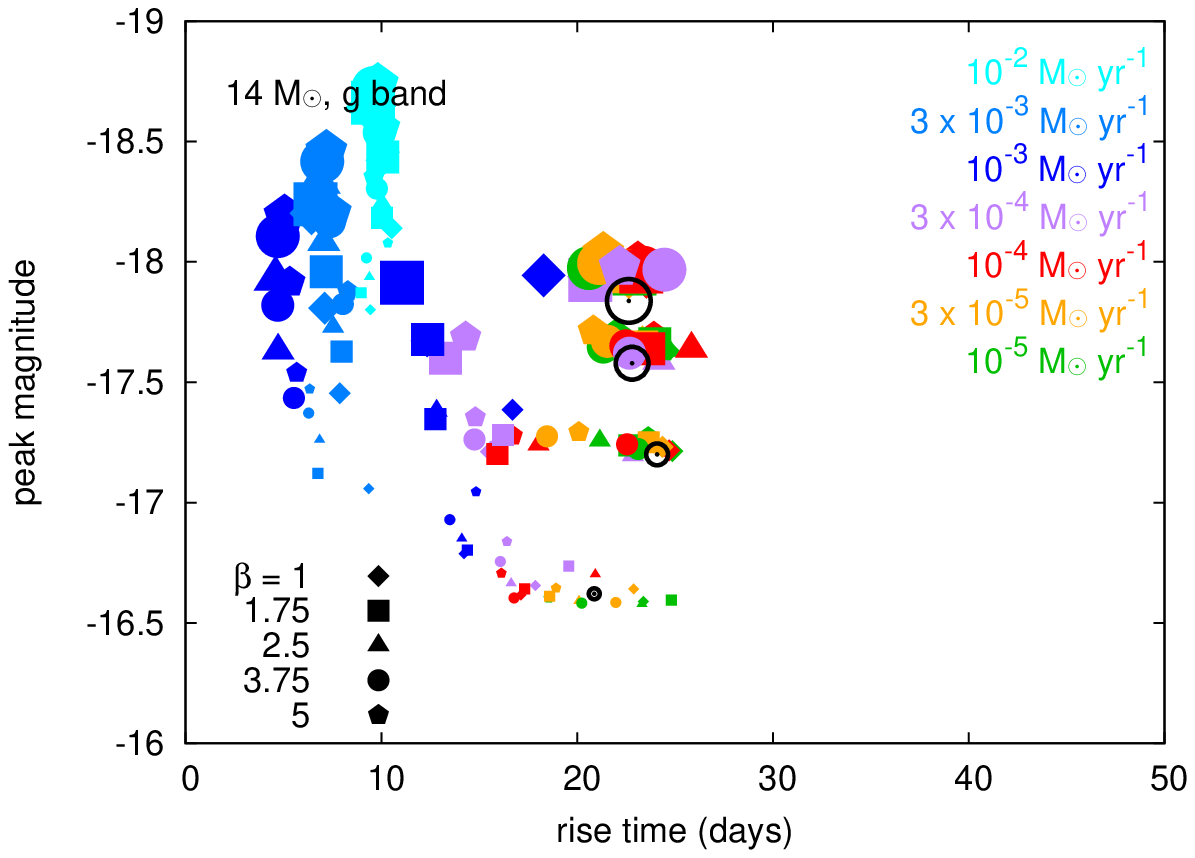} \\
   \includegraphics[width=0.9\columnwidth]{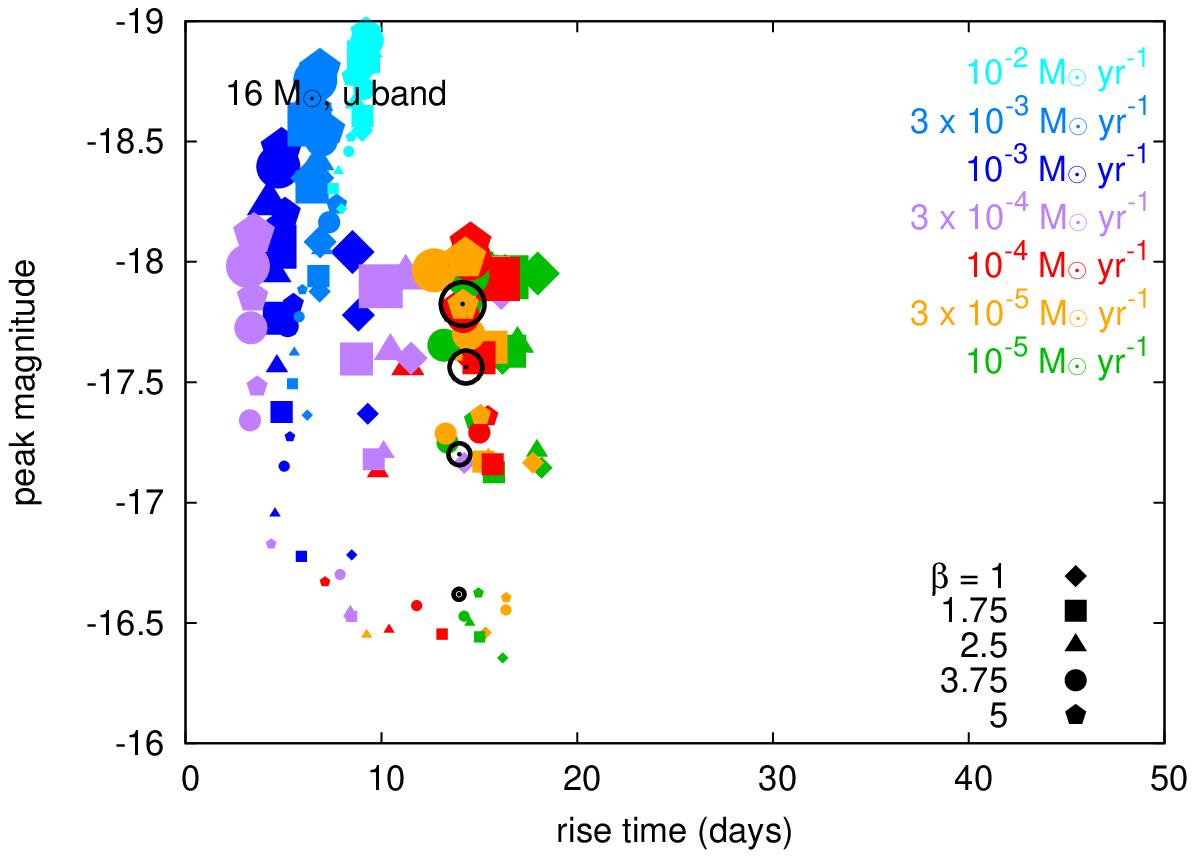}
   \includegraphics[width=0.9\columnwidth]{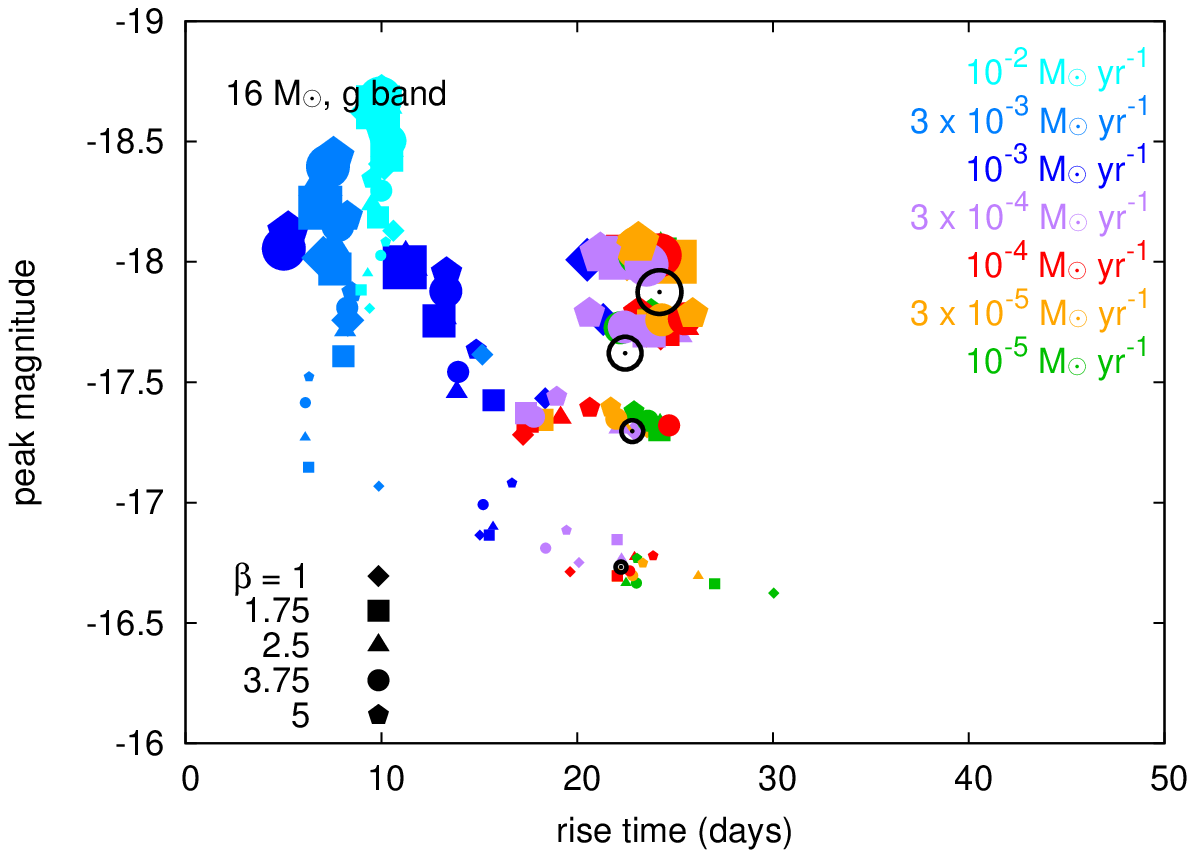}
   \caption{
   Rise times and peak magnitudes of SN~IIP LCs in the $u$ (left panels) and $g$ bands (right panels). The progenitor masses are 12~\Msun\ (top panels), 14~\Msun\ (middle panels), and 16~\Msun\ (bottom panels). The shape and color of the symbols indicate $\beta$ and $\Mdot$, respectively, as shown in the panels. The size of the symbols shows the explosion energies. From the largest  to the smallest symbols, the corresponding explosion energies are $2\times 10^{51}~\mathrm{erg}$, $1.5\times 10^{51}~\mathrm{erg}$, $10^{51}~\mathrm{erg}$, and $5\times 10^{50}~\mathrm{erg}$. Open circles are the models without CSM.
   }
    \label{fig:risepropsdss}
\end{figure*}

\begin{figure*}
   \includegraphics[width=0.9\columnwidth]{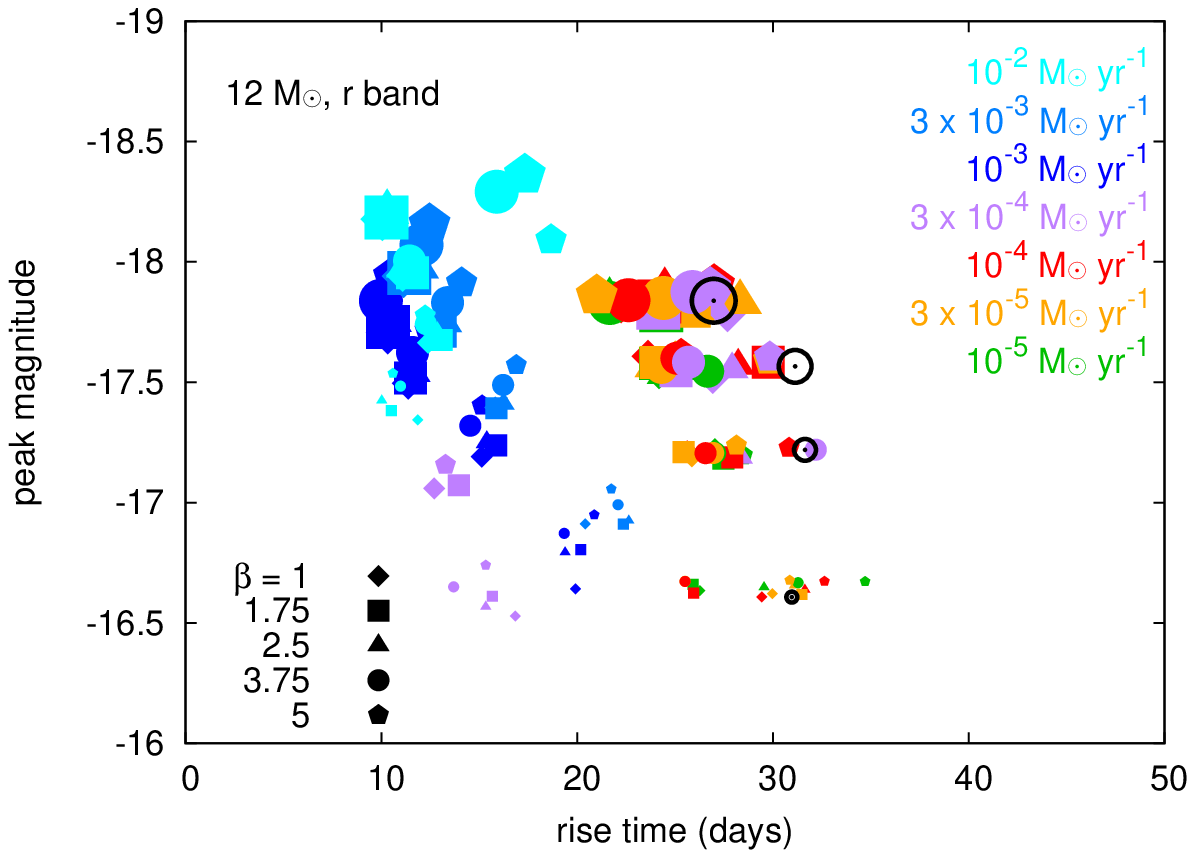}
   \includegraphics[width=0.9\columnwidth]{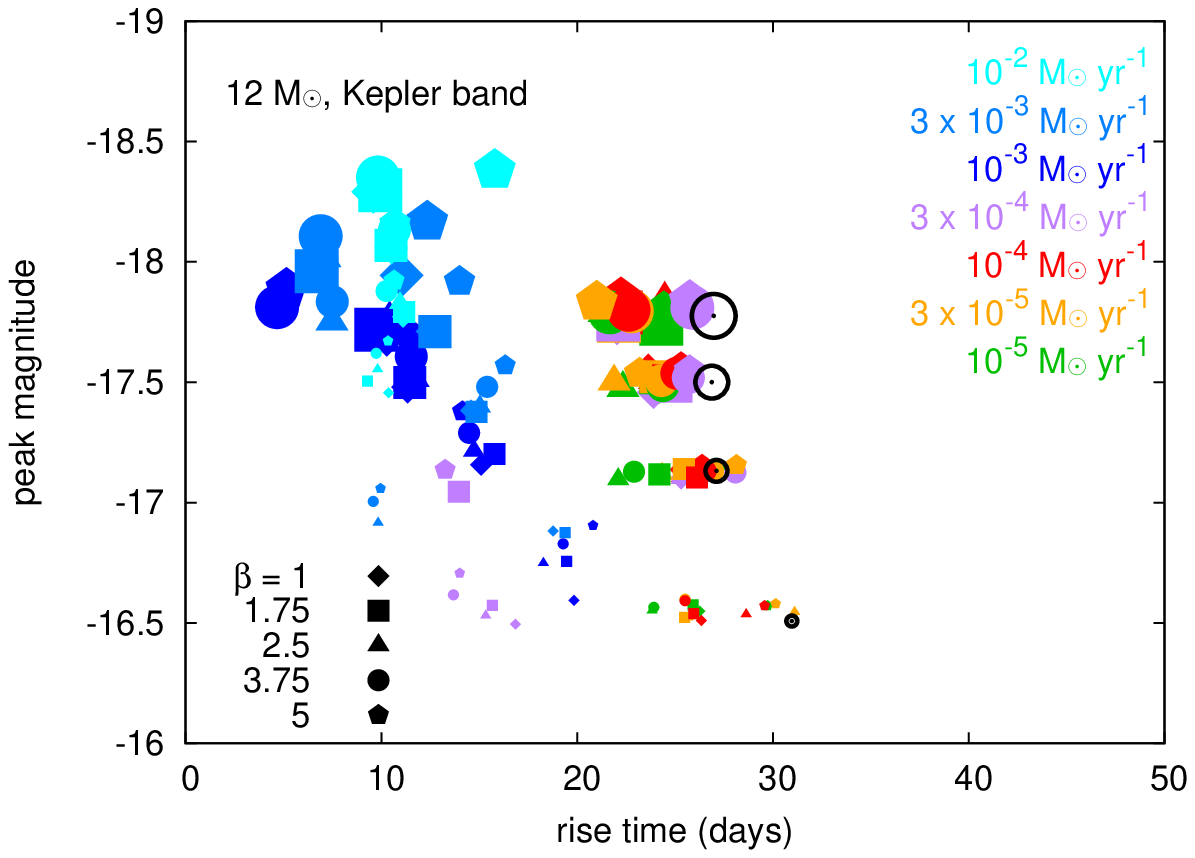} \\
   \includegraphics[width=0.9\columnwidth]{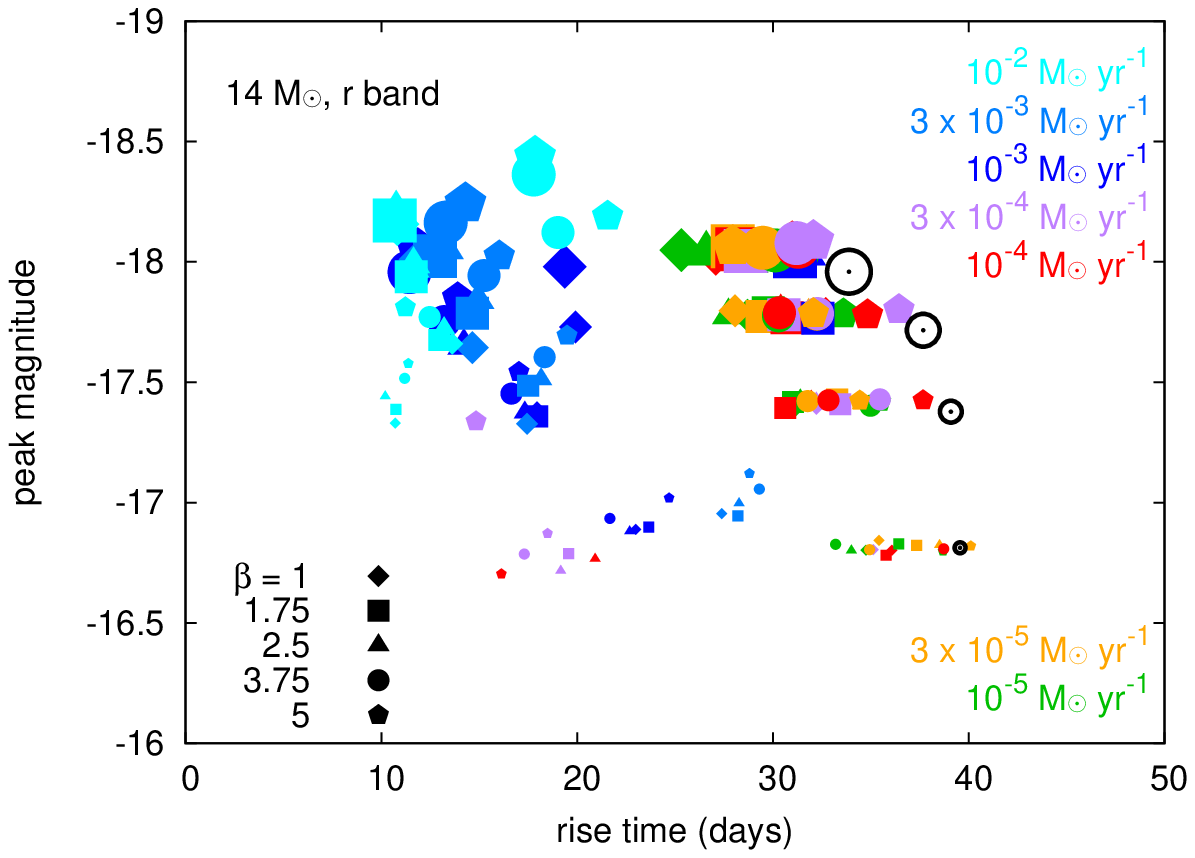}
   \includegraphics[width=0.9\columnwidth]{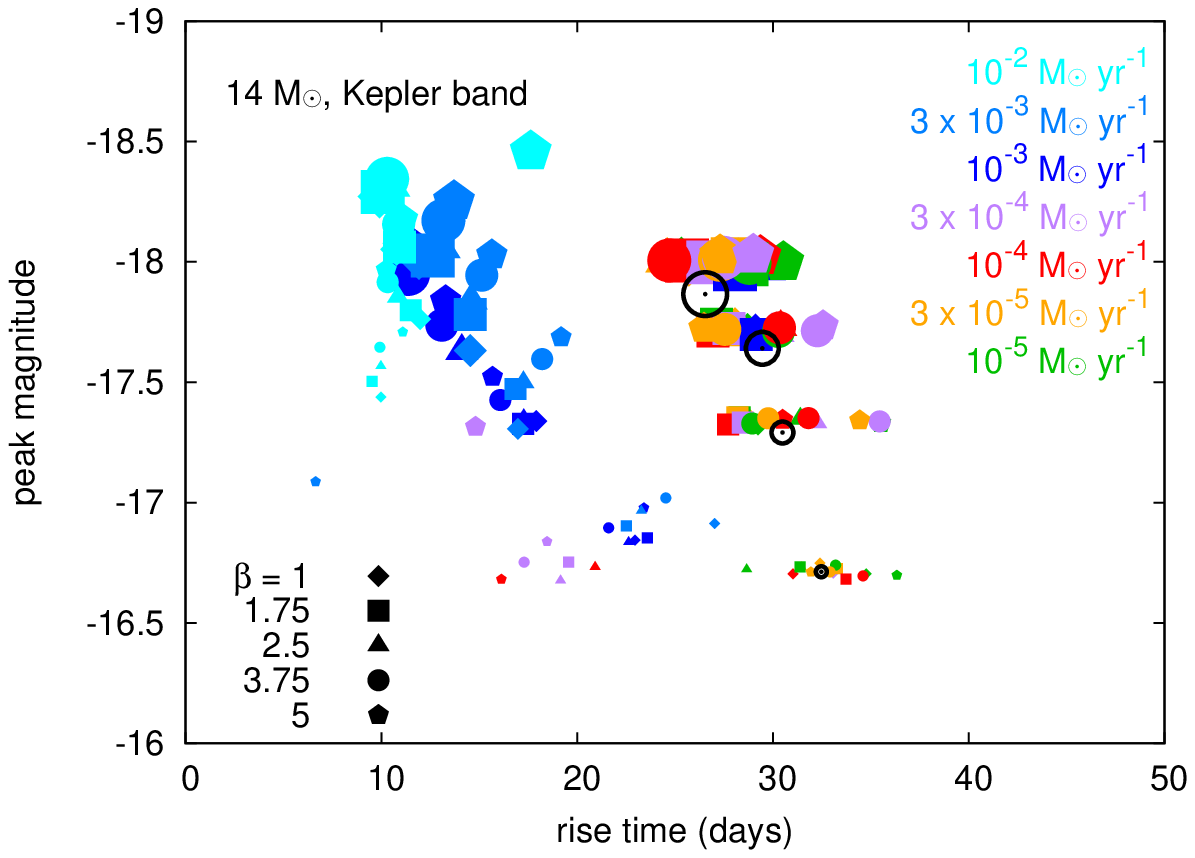} \\
   \includegraphics[width=0.9\columnwidth]{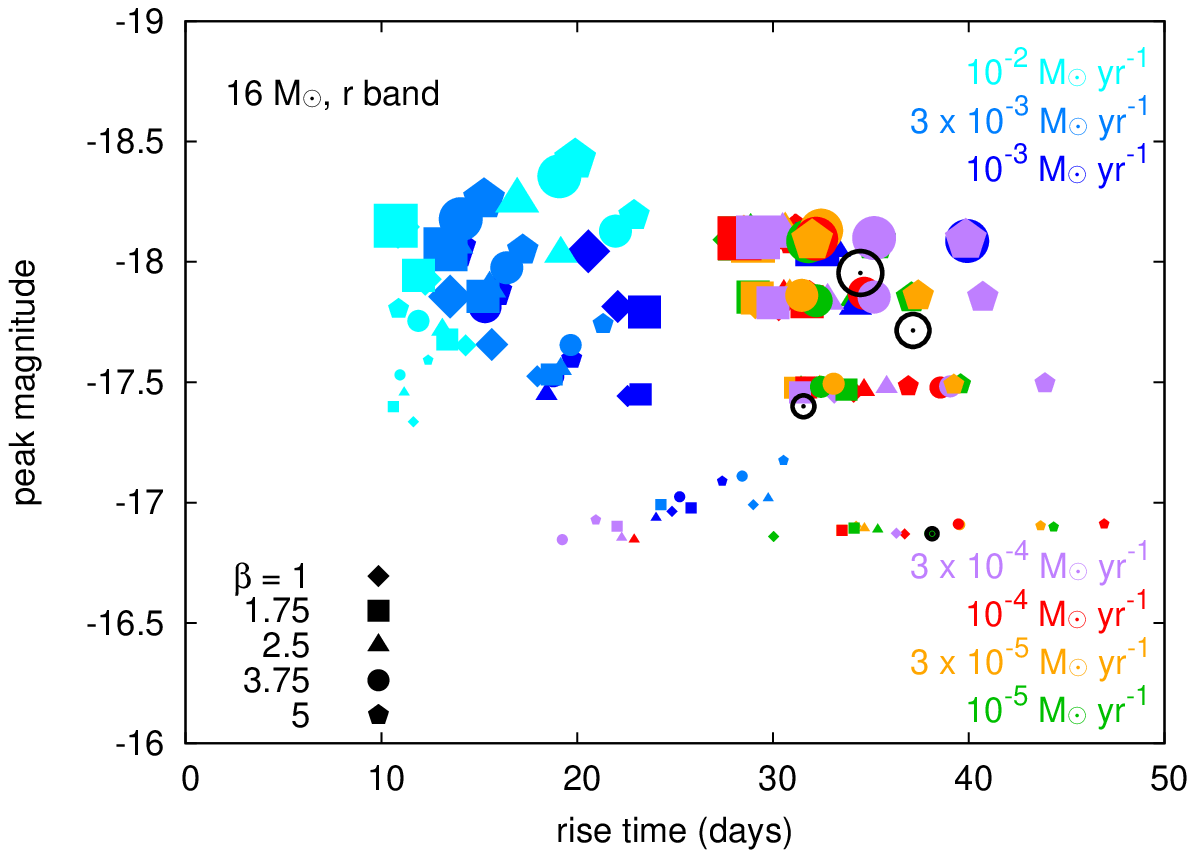}
   \includegraphics[width=0.9\columnwidth]{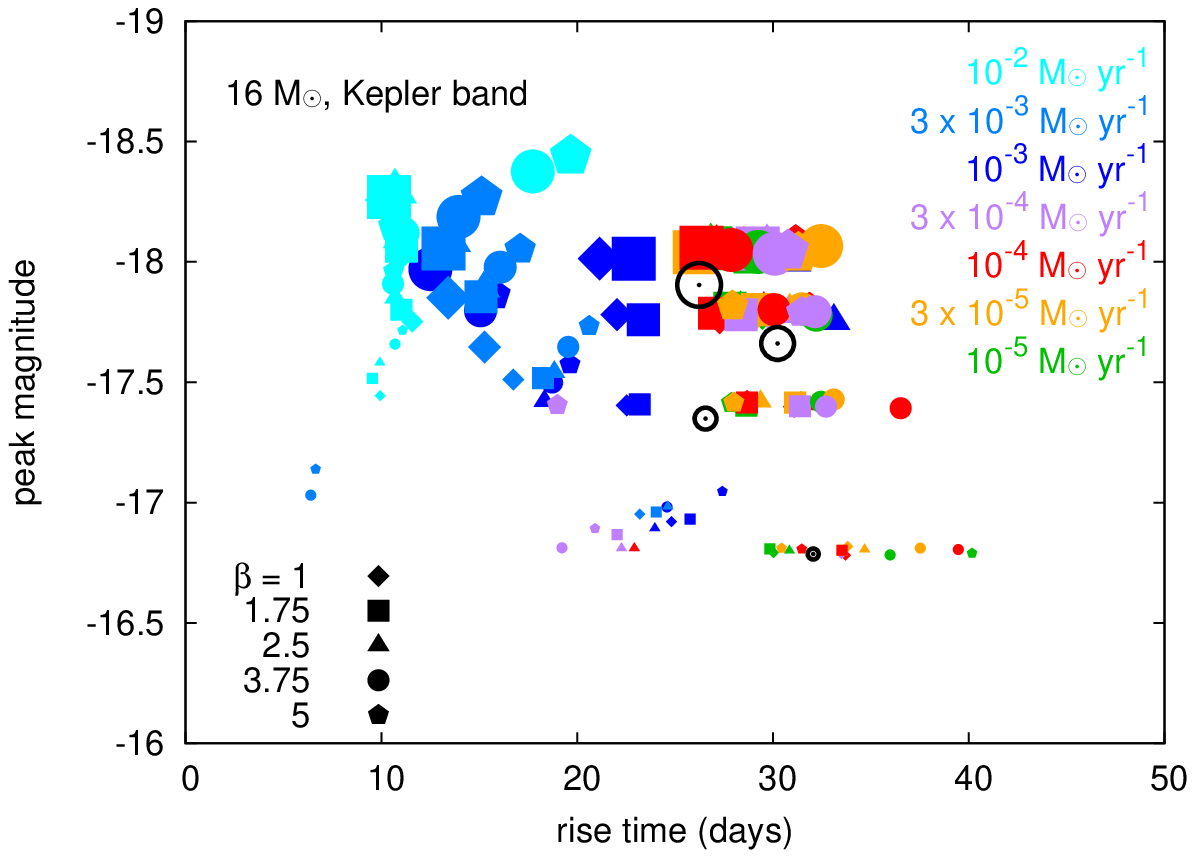}
   \caption{
   Same as Fig.~\ref{fig:risepropsdss} but for the $r$ (left panels) and Kepler (right panels) bands.
   }
    \label{fig:risepropkepler}
\end{figure*}

\subsection{Light curve calculations}\label{sec:lcmethods}
SN LCs presented in this study are all calculated by using the one-dimensional multi-group radiation hydrodynamics code \texttt{STELLA} \citep[e.g.,][]{blinnikov1998sn1993j,blinnikov2000sn1987a,blinnikov2006sniadeflg}, which is also used in our previous studies of RSG explosions within dense CSM \citep{moriya2011iipcsm,moriya2017windacc13fs}. The code follows the time evolution of the spectral energy distribution (SED) and LCs at specific bands can be obtained by convolving the filter functions with the SEDs. We use 100 bins divided in a log scale in the frequency space ranging from 1~\AA\ to 50000~\AA, which is the standard setup of the code.

We adopt 4 different filters to obtain LCs, i.e., the $u$, $g$, $r$, and the Kepler band. The $u$, $g$, and $r$ bands are from the Sloan Digital Sky Survey \citep{doi2010sdss}. The $g$ band is often used in the high-cadence transient surveys. The $u$ and $r$ bands are bluer and redder bands than the $g$ band, respectively. The Kepler satellite has caught the initial rise of a few SNe~IIP and its observation is very dense in time \citep{garnavich2016keplerbreakout}. It has a broad wavelength coverage ranging from around 4000~\AA\ to 9000~\AA\footnote{\url{http://keplergo.arc.nasa.gov/Instrumentation.shtml}}. 

We take the SN progenitors introduced in the previous section and set them as the initial condition in \texttt{STELLA}. We set the mass cut at 1.4~\Msun\ and put thermal energy just above the mass cut to initiate the explosions. All the models are assumed to have 0.1~\Msun\ of \Ni\ at the center but this does not affect the early LCs we focus in this paper. All the SED evolution and LC data calculated for this paper are available at \url{https://goo.gl/o5phYb}.

\section{Results}\label{sec:results}
We present the results of our LC calculations. We base our discussion on the models obtained from our 14~\Msun\ progenitors. We refer to the LCs from the 14~\Msun\ progenitor unless otherwise mentioned.

\subsection{Effect of the wind acceleration on early LCs}
Fig.~\ref{fig:lc_lows} demonstrates the effect of the wind acceleration on the early SN LCs. The left panel shows LCs with a fixed $\beta=1$ and different mass-loss rates. The right panel shows the LCs with a fixed $\Mdot=10^{-5}~\Msunpyr$ and different $\beta$. We can find that the $g$-band LCs with $\Mdot\simeq (1-3)\times 10^{-4}~\Msunpyr$ can be reproduced by the CSM with $\Mdot=10^{-5}~\Msunpyr$ and $\beta\gtrsim 2.5$ without increasing the mass-loss rates. The large $\beta$ increases the CSM density near the progenitors and has the same effect as the high mass-loss rates (Fig.~\ref{fig:density}). Indeed, the CSM mass in the model with $3\times10^{-4}~\Msunpyr$ and $\beta=1$ is 0.018~\Msun\ and it is similar to that of the model with $10^{-5}~\Msunpyr$ and $\beta=2.5$ (0.013~\Msun). 

If the CSM density continues to follow $\rhocsm\propto r^{-2}$ up to the stellar surface, the CSM mass within $10^{15}~\mathrm{cm}$ is $3\times 10^{-4}~\Msun$ if $\Mdot=10^{-5}~\Msunpyr$ and $\vwind=10~\kmps$. However, even with the conservative assumption of $\beta=1$, the CSM mass becomes $3\times 10^{-3}~\Msun$ which is already a factor of 10 larger. This demonstrates the importance of taking the wind acceleration into account in properly estimating the mass-loss properties of the SN progenitors. 

The $10^{-5}~\Msunpyr$ model with $\beta=5$ has a CSM mass of 0.210~\Msun. This CSM mass is comparable to that of the $3\times 10^{-3}~\Msunpyr$ model with $\beta=1$ ($0.137~\Msun$). However, the $g$-band luminosity of the $10^{-5}~\Msunpyr$ model is much fainter than the $g$-band luminosity of the $3\times 10^{-3}~\Msunpyr$ model (Fig.~\ref{fig:lc_lows}). If we see the bolometric LCs of the models in Fig.~\ref{fig:lc_lows}, we can find that the $10^{-5}~\Msunpyr$ model actually gets brighter than the $10^{-3}~\Msunpyr$ model, although the duration of the bright phase is shorter. This is because the CSM mass is more concentrated near the progenitor in the $10^{-5}~\Msunpyr$ model with $\beta = 5$ than in the $3\times 10^{-3}~\Msunpyr$ model with $\beta = 1$ (cf. Fig.~\ref{fig:density}). Thus, the interaction occurs intensively in a shorter period in the former model, making the interaction phase shorter and brighter. The diffusion time in the CSM is also larger in the $10^{-3}~\Msunpyr$ model having the higher density in the outer layers, making the duration of the bright phase long. The concentrated CSM with less diffusion also results in a hotter photosphere which makes the luminosity increase in the $g$ band less significant.

Some $g$-band LCs in Fig.~\ref{fig:lc_lows} have a separate initial LC peak at several days after the explosion. They can also be understood by the existence of the concentrated CSM. The initial LC peak only appears in the models where the CSM mass is less than $\sim 0.01~\Msun$. The concentrated dense CSM with a relatively small mass can only affect the LCs at the very beginning making the initial LC peak and they immediately go back to the same LC as the one without CSM. This kind of initial peak resembles the early peak of a stripped-envelope SN from a progenitor having an extended envelope \citep[e.g.,][]{piro2015doublepeak}.

The existence of the dense CSM just above the progenitor allows the early LCs to rise much faster than those without CSM. Even in the case of $10^{-5}~\Msunpyr$, in which no significant effect on SN LCs is found in previous studies \citep[e.g.,][]{moriya2011iipcsm}, we find a quick rise of the LC. This is because of the higher density in the immediate stellar vicinity caused by the wind acceleration (Table~\ref{tab:csmmass} and Fig.~\ref{fig:density}). However, most of the CSM mass is concentrated only near the stellar surface and the dense CSM is quickly swept by the shock. Therefore, we do not see a continuous effect on the LC and the luminosity soon goes back to that of the model without CSM, except for the extreme case of $\Mdot\gtrsim 10^{-2}~\Msunpyr$ \citep[cf.][]{morozova2017iil}.

\subsection{Multicolor LCs and color evolution}
We have demonstrated that a large $\beta$ for a given mass-loss rate has a similar effect to a high mass-loss rate for a given $\beta$ on the early SN~IIP LCs. We here show their observational filter dependences.

The left panels of Fig.~\ref{fig:lc_multiclc} show the LCs with a constant $\Mdot = 10^{-3}~\Msunpyr$ but different $\beta$ ranging from $\beta=1$ to 5. The right panels of Fig.~\ref{fig:lc_multiclc} show the LCs with a constant $\beta=3.75$ but different $\Mdot$ ranging from $10^{-5}~\Msunpyr$ to $10^{-3}~\Msunpyr$. As found in the previous section, the increase in $\beta$ has similar effects to the increase in \Mdot. The LCs in the bluer bands rise more sharply than in the redder bands. The Kepler band LCs behave in a similar way to the $r$-band LCs.

Fig.~\ref{fig:color} shows representative color ($g-r$) evolution of our LC models. The overall color evolution is similar in the interaction models but the timescales of the evolution depend on the CSM configuration. The color evolution becomes similar to those of the models without CSM when the interaction ends, except for the models with very massive CSM ($\sim 1~\Msun$ or more) as in the $10^{-2}~\Msunpyr$ model (see also Section~\ref{sec:longterm}).

\begin{figure*}
   \includegraphics[width=0.9\columnwidth]{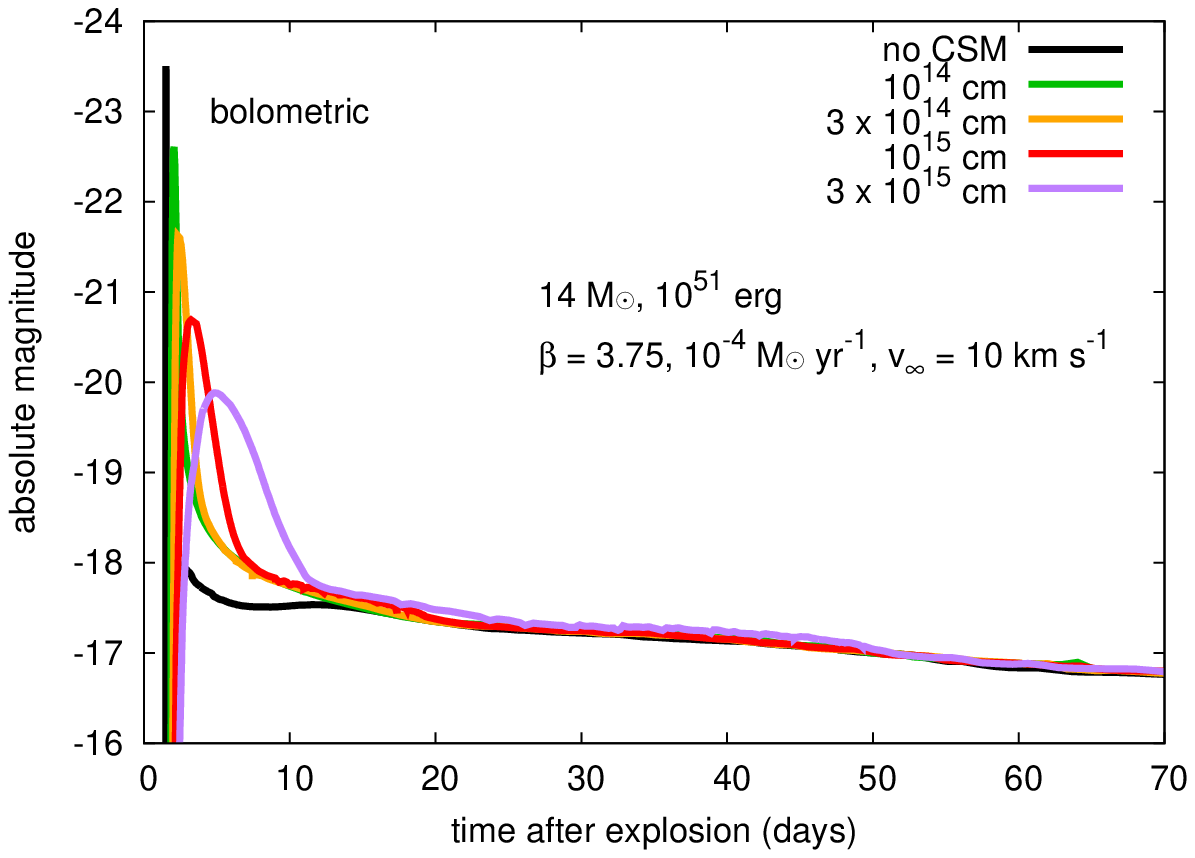}
   \includegraphics[width=0.9\columnwidth]{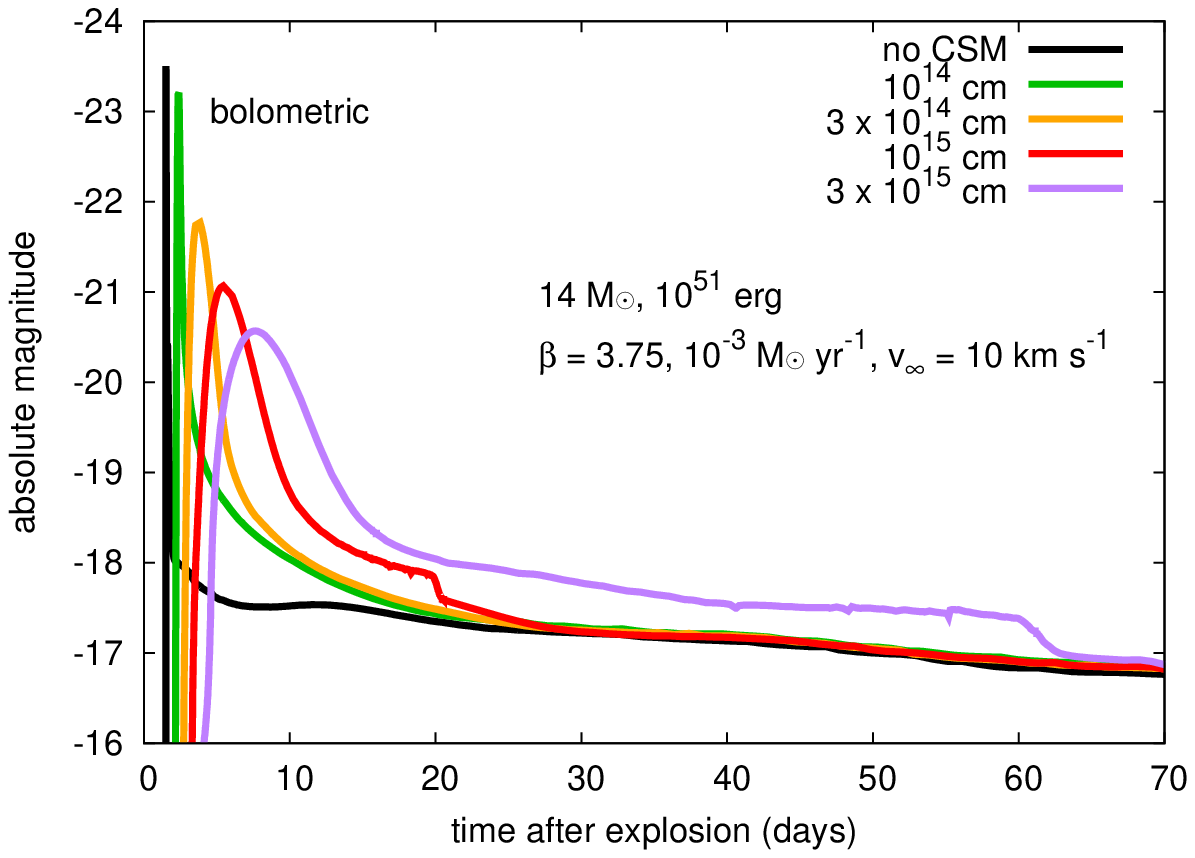}  \\
   \includegraphics[width=0.9\columnwidth]{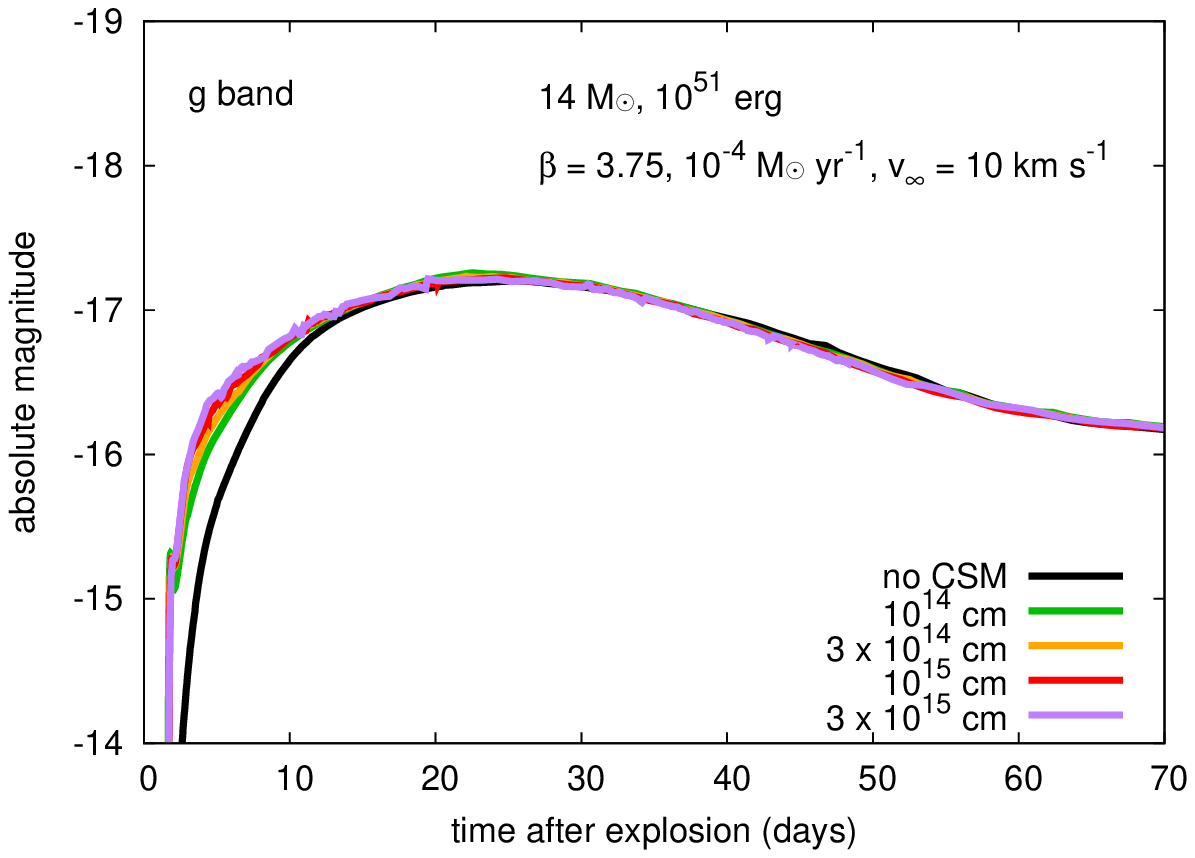}
   \includegraphics[width=0.9\columnwidth]{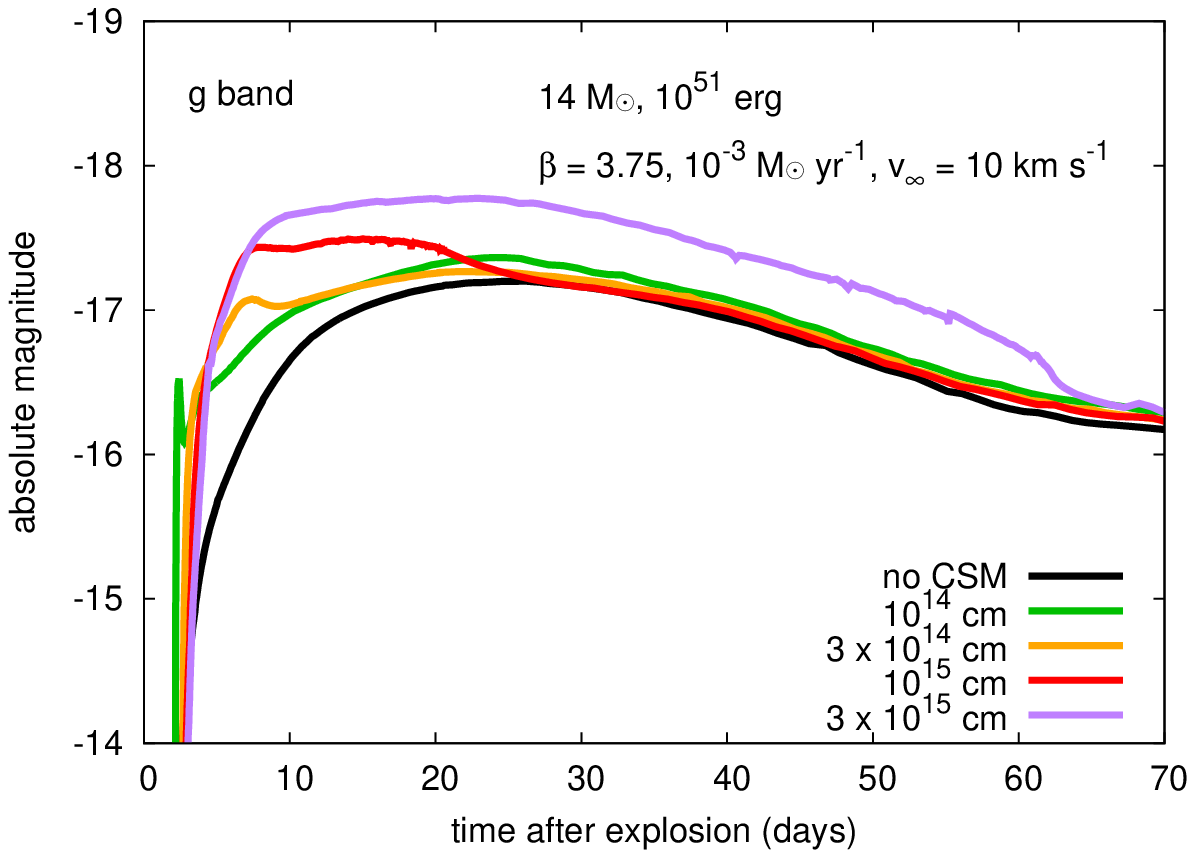}   
   \caption{
   LCs with different dense CSM radii. The left panels show the models with $\Mdot=10^{-4}~\Msunpyr$ and the right panels show the models with $\Mdot=10^{-3}~\Msunpyr$.  All the models have the 14~\Msun\ progenitor, $\beta=3.75$, and the explosion energy of $10^{51}~\mathrm{erg}$.
   }
    \label{fig:lc_diffrad}
\end{figure*}

\begin{figure*}
   \includegraphics[width=0.9\columnwidth]{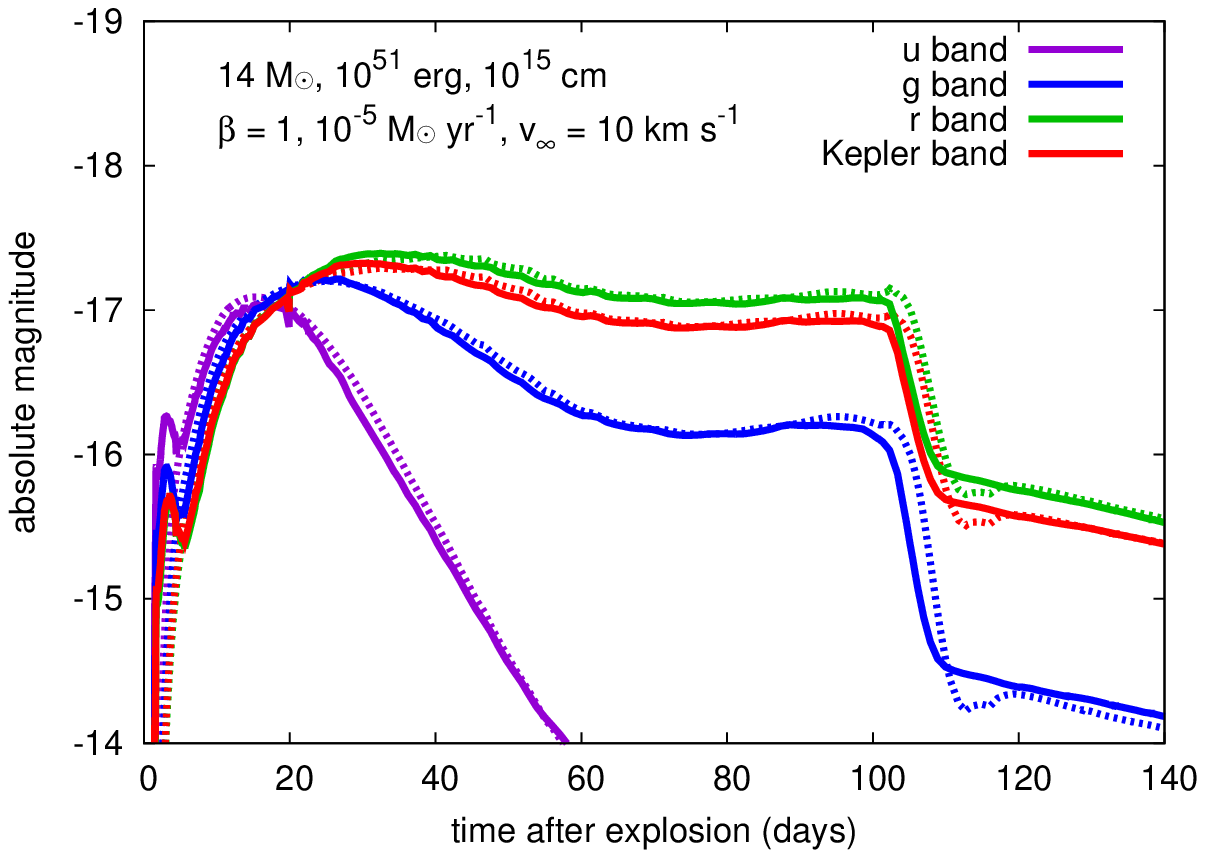}
   \includegraphics[width=0.9\columnwidth]{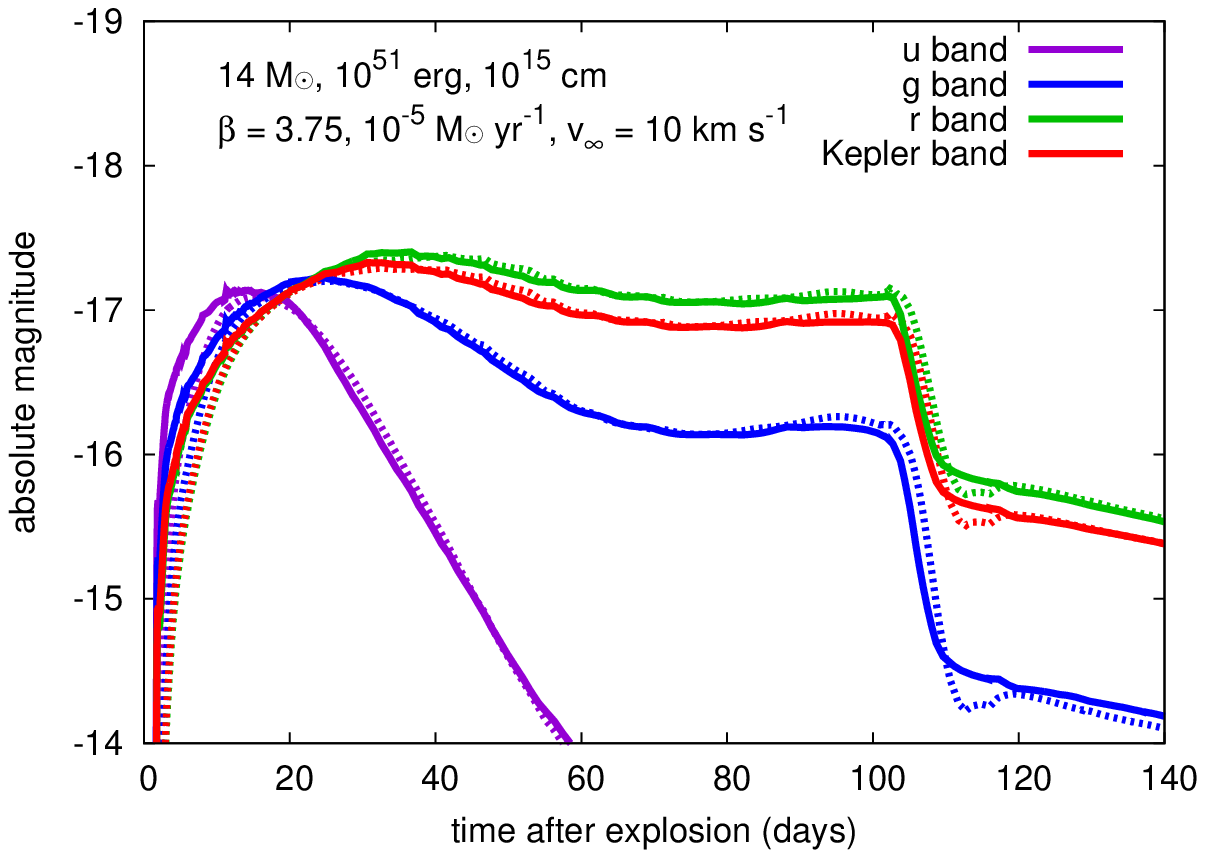} \\
   \includegraphics[width=0.9\columnwidth]{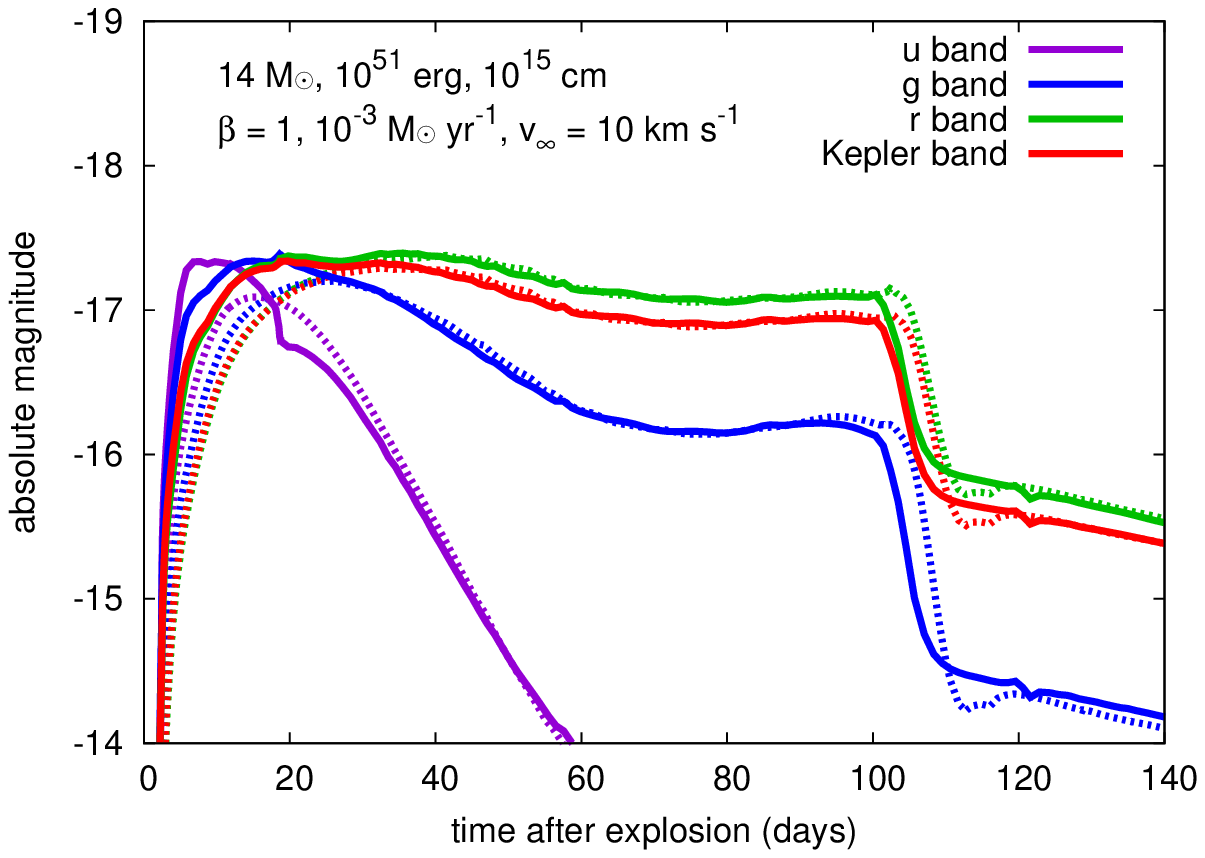}
   \includegraphics[width=0.9\columnwidth]{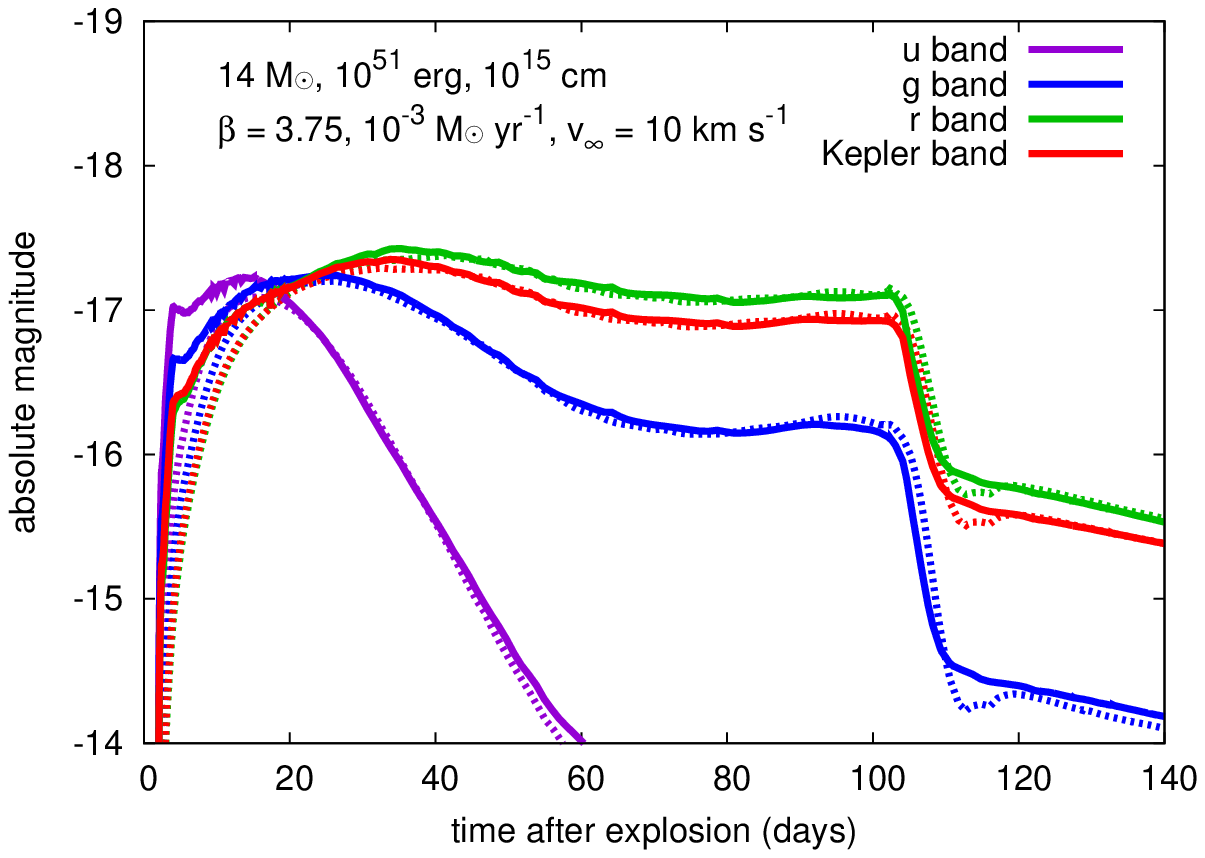} \\
   \includegraphics[width=0.9\columnwidth]{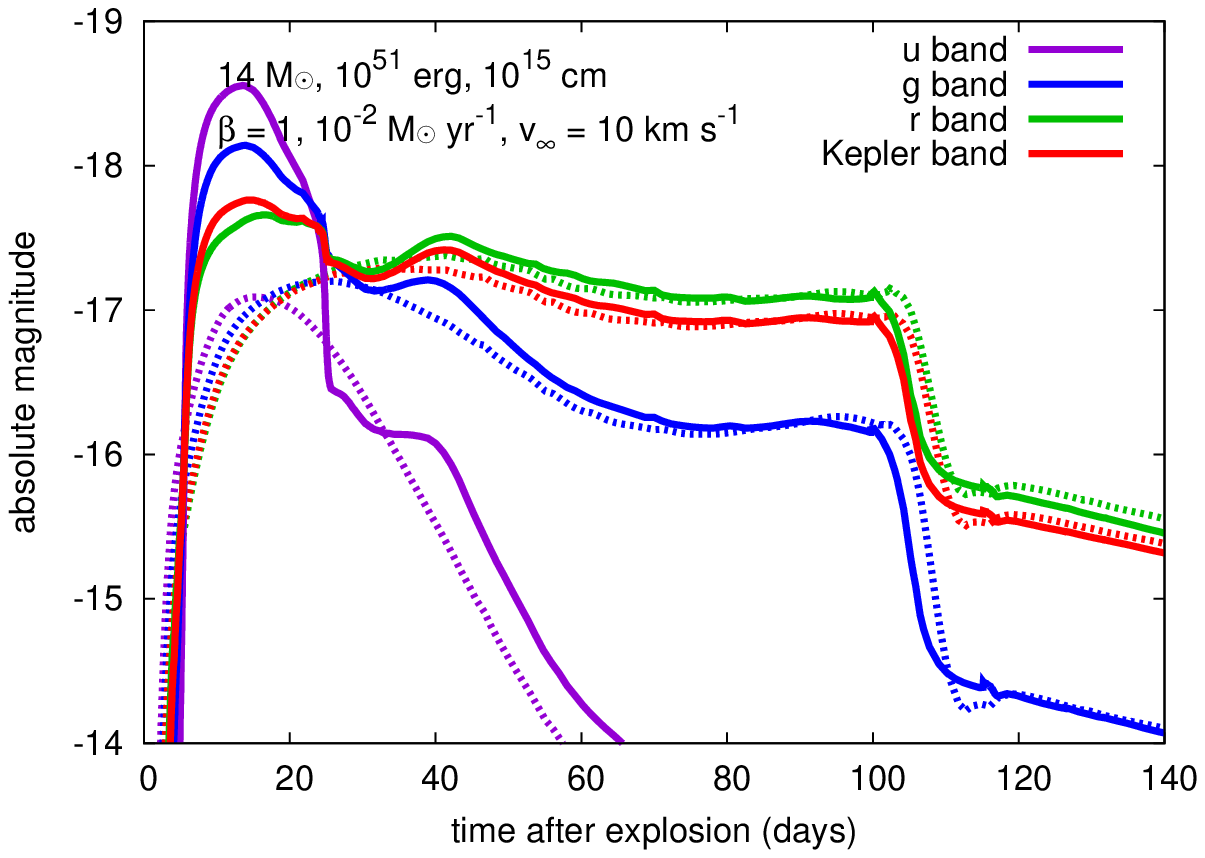}
   \includegraphics[width=0.9\columnwidth]{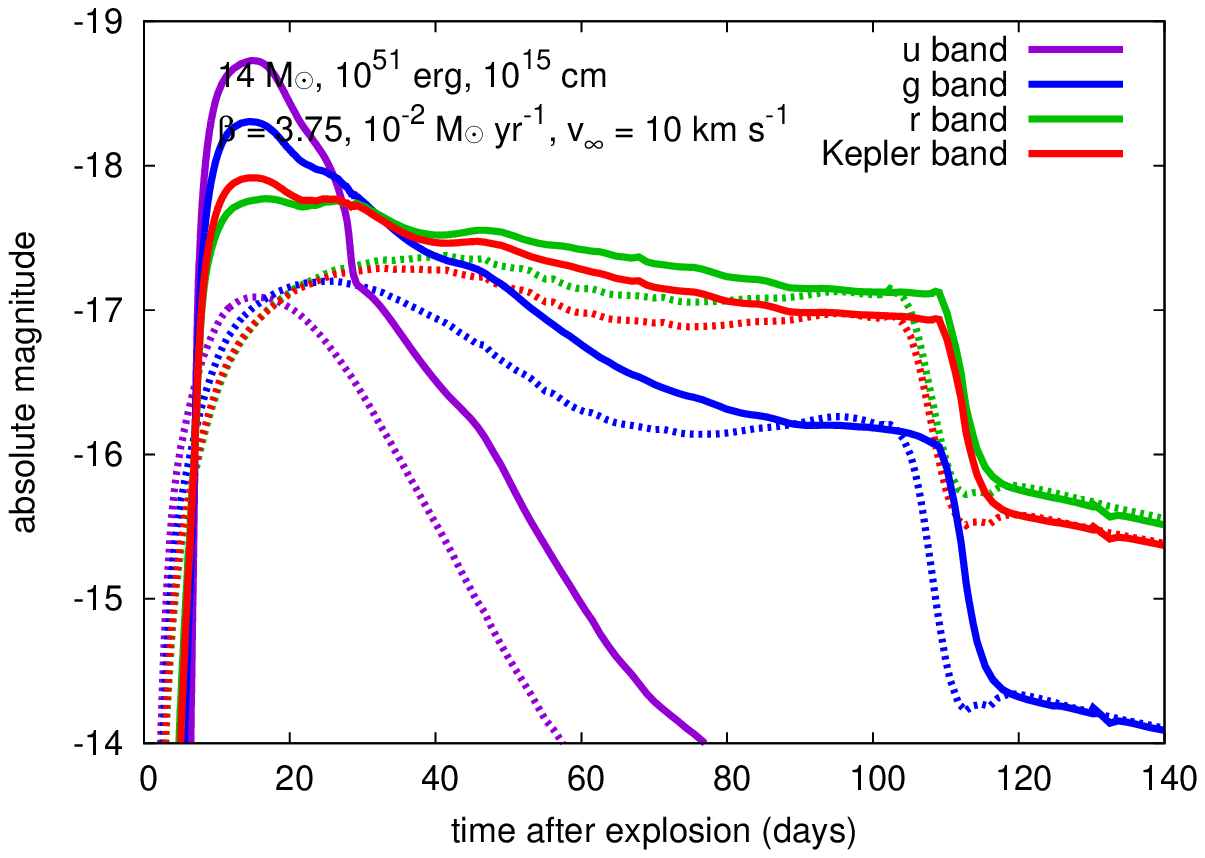}
   \caption{
   Long-term multicolor LCs. The dotted LCs are without dense CSM.
   }
    \label{fig:multicolorlightcurve}
\end{figure*}

\subsection{Rise times}\label{sec:risetimes}
For all the models with different progenitors, explosion energies, and CSM properties, we measure the rise time of the LCs in the $u$, $g$, $r$, and Kepler bands. The rise times in the $u$ and $g$ bands are summarized in Fig.~\ref{fig:risepropsdss} and those in the $r$ and Kepler bands are summarized in Fig.~\ref{fig:risepropkepler}.

We measure the rise time as the required time for a LC to reach the peak magnitude from $-13$~mag in the given band. $-13$~mag is arbitrarily chosen. However, the rise times are not necessarily a good indicator of the LC rising rate. For example, the LCs with dense CSM in Fig.~\ref{fig:lc_lows} have much faster rise at the beginning than those without CSM. However, after the initial quick rise, some LCs with dense CSM flatten and the time required to reach the maximum brightness is not much different from the LC without CSM. Therefore, although the initial LC rising rates are different due to the existence of the CSM, the rise time can still be similar in the LCs with and without CSM.

One can find in Figs.~\ref{fig:risepropsdss} and \ref{fig:risepropkepler} that the peak magnitudes tend to be brighter with larger $\beta$ for a given mass-loss rate. This is because the CSM mass to decelerate the SN ejecta is larger in the CSM with a larger $\beta$. When \Mdot\ is low, the LCs with low \Mdot\ with large $\beta$ tend to have similar rise times to those with high \Mdot\ with small $\beta$. However, when \Mdot\ is sufficiently high ($\Mdot\gtrsim 10^{-3}~\Msunpyr$), the rise times are mainly determined by the mass-loss rates. This is because the radius where the electron-scattering optical depth becomes unity is well beyond the radius where the wind acceleration is taking place regardless of $\beta$. Therefore, the diffusion time in the CSM is similar regardless of $\beta$ in the high \Mdot\ models. The rise times are particularly well separated by the mass-loss rates in the $g$ band.

The rise times become shorter as \Mdot\ increases up to a certain \Mdot\ but then they start to be longer as \Mdot\ increases. This is because of the increase in the diffusion time in the dense CSM caused by the increase in density. Up to $\Mdot\simeq 10^{-4}~\Msunpyr$, photons released due to the interaction do not diffuse much in the CSM because of the relatively small outer CSM density. However, with a higher \Mdot, photons from the interaction start to diffuse in the unshocked CSM and the rise time starts to increase. The  shock breakout also occurs in the CSM when the CSM density becomes large enough \citep[e.g.,][]{ofek2010ptf09uj,chevalier2011irwin,moriya2011iipcsm}.

The rise times are shorter in the bluer band. The rise times without CSM are around $10-15$~days in the $u$ band, around $20$~days in the $g$ band, and around $20-40$~days in the $r$ band. The rise times are not strongly altered when the mass-loss rates are less than $\sim 3\times 10^{-4}~\Msunpyr$. They get significantly shorter (less than $\sim 10~\mathrm{days}$ in the $u$ and $g$ bands and less than $\sim 20~\mathrm{days}$ in the $r$ band) when the mass-loss rates are more than $\sim 3\times 10^{-4}~\Msunpyr$. 
The rise times in the Kepler band behave similar to those in the $r$ band.

\subsection{Effect of CSM radii}\label{sec:radii}
All the models presented so far have a CSM radius of $10^{15}~\mathrm{cm}$. The radius of the dense CSM affects the duration of the interaction between SN ejecta and dense CSM and, therefore, the duration of the time for which we can observe the interaction signatures in the SN. In the case of SN~2013fs, the interaction signatures disappeared in about 5~days after the explosion and the dense CSM is suggested to be confined within $10^{15}~\mathrm{cm}$ \citep{yaron2017iipcsm}. When \Mdot\ is sufficiently high, the CSM radius also affects the diffusion time in the CSM because the radius of the optically thick CSM can be larger in the more extended CSM. Here, we show LC models with several different dense CSM radii ($10^{14}~\mathrm{cm}$, $3\times 10^{14}~\mathrm{cm}$, $10^{15}~\mathrm{cm}$, and $3\times 10^{15}~\mathrm{cm}$) from the two different mass-loss rates ($10^{-4}~\Msunpyr$ and $10^{-3}~\Msunpyr$). We keep all the other parameters fixed to: a progenitor mass of 14~\Msun, an explosion energy of $10^{51}~\mathrm{erg}$, and $\beta=3.75$.

Although the CSM radii vary by a factor of 30, the CSM mass in the models are not much different from each other. The CSM masses in the $10^{-3}~\Msunpyr$ model are 0.403~\Msun\ ($10^{14}~\mathrm{cm}$), 0.430~\Msun\ ($3\times 10^{14}~\mathrm{cm}$), 0.459~\Msun\ ($10^{15}~\mathrm{cm}$), and 0.525~\Msun\ ($3\times 10^{15}~\mathrm{cm}$) and those in the $10^{-4}~\Msunpyr$ model are 0.195~\Msun\ ($10^{14}~\mathrm{cm}$), 0.195~\Msun\ ($3\times 10^{14}~\mathrm{cm}$), 0.195~\Msun\ ($10^{15}~\mathrm{cm}$), and 0.199~\Msun\ ($3\times 10^{15}~\mathrm{cm}$). This is because most of the mass is concentrated within $10^{14}~\mathrm{cm}$ (cf. Fig.~\ref{fig:density}). However, the radius of the optically thick region in the CSM is affected by the CSM radius when the mass-loss rate is high enough. Assuming the electron-scattering opacity of $0.34~\mathrm{cm^2~g^{-1}}$, the photosphere where the optical depths become 2/3 is located at $10^{14}~\mathrm{cm}$ (the $10^{14}~\mathrm{cm}$ model), $3\times 10^{14}~\mathrm{cm}$ (the $3\times 10^{14}~\mathrm{cm}$ model), $8\times 10^{14}~\mathrm{cm}$ (the $10^{15}~\mathrm{cm}$ model), and $1.5\times 10^{15}~\mathrm{cm}$ (the $3\times 10^{15}~\mathrm{cm}$ model) in the $10^{-3}~\Msunpyr$ model. On the contrary, the location of the photosphere is not strongly affected by the CSM radii in the $10^{-4}~\Msunpyr$ model, i.e., $10^{14}~\mathrm{cm}$ (the $10^{14}~\mathrm{cm}$ model), $2\times 10^{14}~\mathrm{cm}$ (the $3\times 10^{14}~\mathrm{cm}$ model), $3\times 10^{14}~\mathrm{cm}$ (the $10^{15}~\mathrm{cm}$ model), and $3\times 10^{14}~\mathrm{cm}$ (the $3\times 10^{15}~\mathrm{cm}$ model).

Fig.~\ref{fig:lc_diffrad} shows the LCs with different CSM radii. The CSM radii have significant effects on the LCs only when \Mdot\ is relatively high. When \Mdot\ is low, the outer CSM density is low and the location of the photosphere does not depend on \Mdot. For example, in the case of $\Mdot=10^{-4}~\Msunpyr$ with $\beta=3.75$, the photosphere is always at $3\times 10^{14}~\mathrm{cm}$ when the CSM radius is larger than $\simeq 3\times 10^{14}~\mathrm{cm}$. Therefore, extending the CSM radius does not change the LCs significantly in the low \Mdot\ cases (Fig.~\ref{fig:lc_diffrad}).

The effect of the CSM radii is relatively large when the mass-loss rates are $\gtrsim 10^{-3}~\Msunpyr$. When the CSM radius is small, the interaction terminates quickly. However, we still find the initial quick rise in the LCs. For $10^{-3}~\Msunpyr$, the models with the larger CSM radii tend to be brighter in the $g$ band but they also have longer rise times. The period of strong interaction is also extended in the more extended CSM models. For example, the interaction in the $10^{15}~\mathrm{cm}$ model terminates in 20~days while the interaction in the $3\times 10^{15}~\mathrm{cm}$ model continues until 60~days.

Contrary to the $g$ band LCs, the bolometric LCs become fainter as the CSM radii become larger. This is because of the larger electron-scattering optical depths caused by the more extended CSM. The larger electron-scattering optical depths widen the bolometric LCs and make the bolometric luminosity smaller because the total radiated energy is more-or-less similar in all the models that contain similar CSM masses. The $g$ band luminosity increase is mainly caused by the higher temperature in the interacting models.

\subsection{Long-term effect}\label{sec:longterm}
We have focused on the early time LCs so far. Here we discuss the long-term LC evolution. Representative long-term multicolor LCs are presented in Fig.~\ref{fig:multicolorlightcurve}. Overall, our dense CSM located within $10^{15}~\mathrm{cm}$ only have strong effects when the LCs are rising and little effect is found in the later phase. This strengthens the necessity of the early SN observations to study the immediate circumstellar environments of the RSG SN progenitors. Long-term effects on the LCs are found in the models with very massive CSM that have $\sim 1~\Msun$ or more \citep[cf.][]{morozova2017iil}.

\begin{figure*}
   \includegraphics[width=0.9\columnwidth]{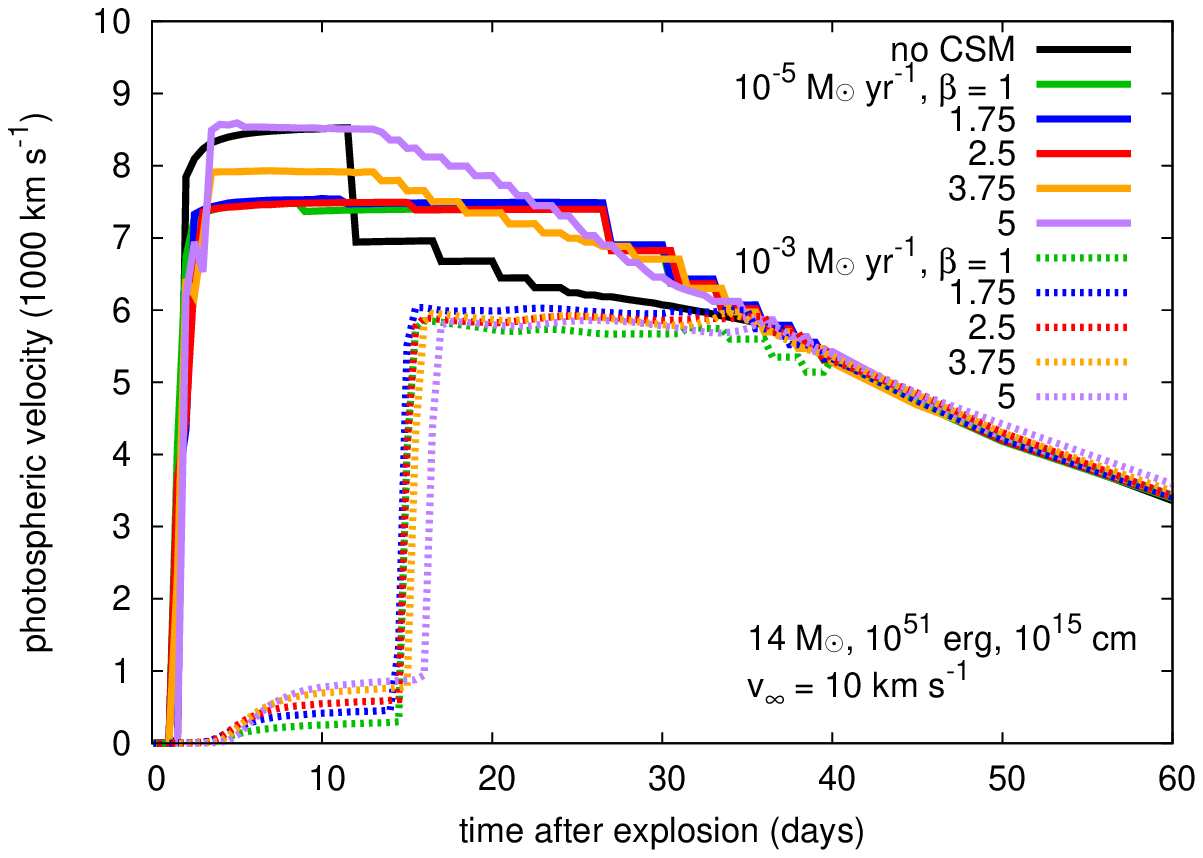}
   \includegraphics[width=0.9\columnwidth]{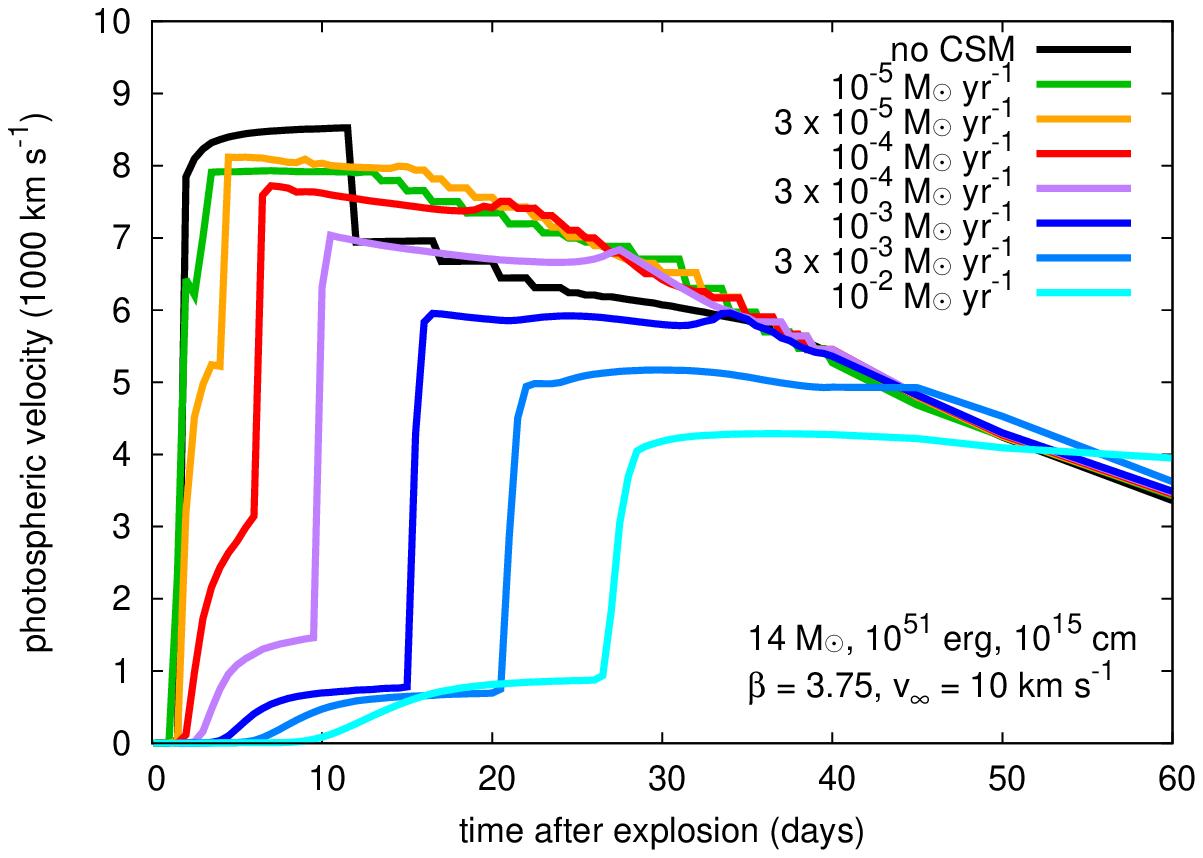} \\
   \includegraphics[width=0.9\columnwidth]{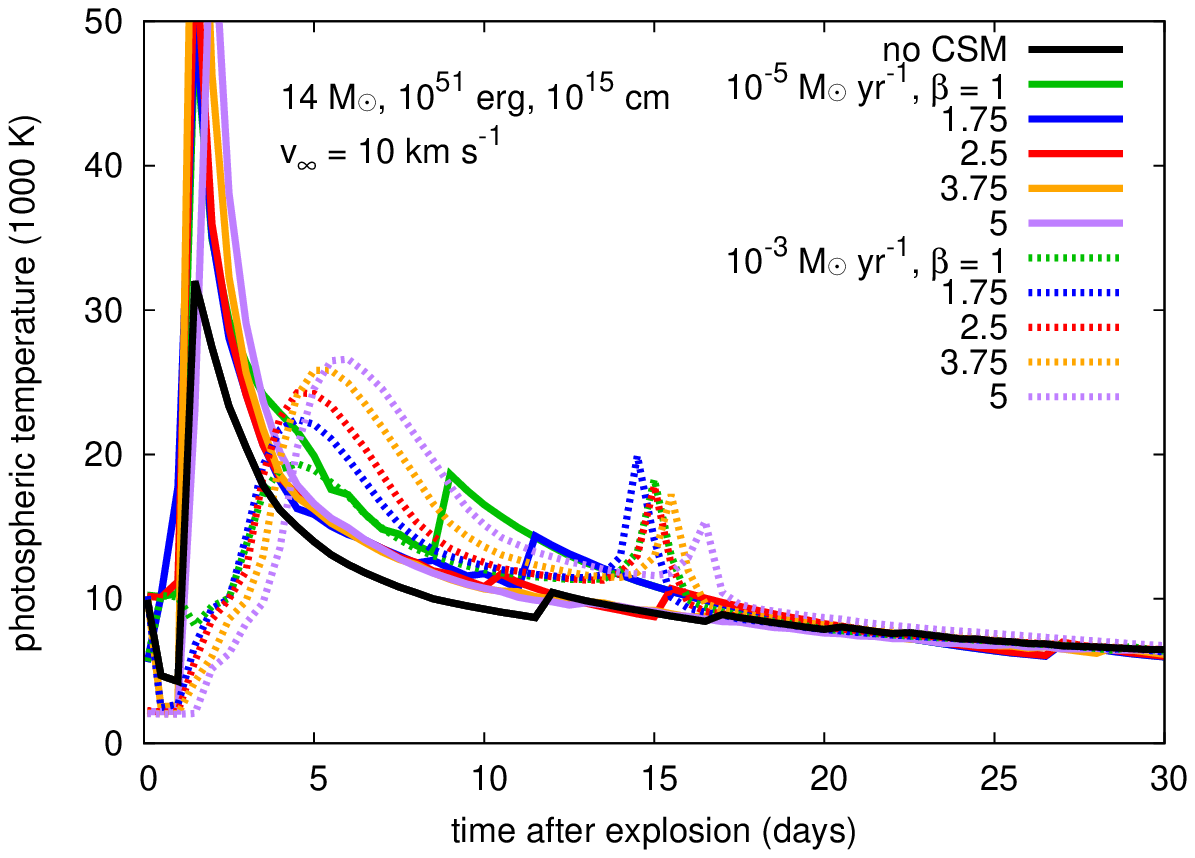}
   \includegraphics[width=0.9\columnwidth]{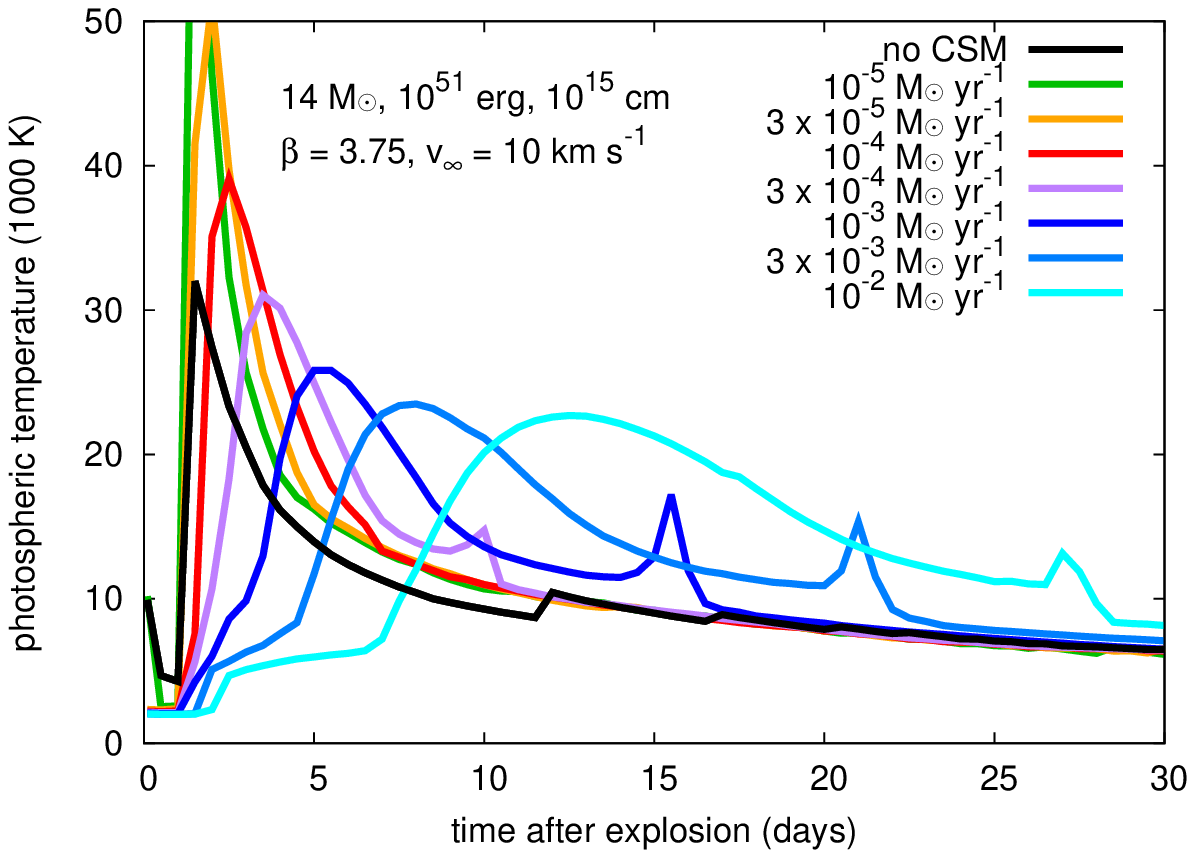}
   \caption{
   Photospheric velocities (top) and temperatures (bottom) of representative models. Note the different time ranges in the top and bottom panels.
   }
    \label{fig:photo}
\end{figure*}

\subsection{Photospheric velocities and temperature}\label{sec:photosphere}
Although our focus in this paper is on the early SN~IIP LC properties, hydrodynamical information is helpful to understand the circumstellar properties of SNe~IIP. Here we present photospheric velocities and temperatures of our models. The photosphere is defined as the location where the Rosseland-mean optical depth becomes 2/3.

The top panels in Fig.~\ref{fig:photo} presents photospheric velocities of the representative models. The evolution of the photospheric velocities are found to be mostly affected by the mass-loss rates, but we do find small effects of the wind acceleration on the photospheric velocities. When the CSM is dense enough, the dense CSM is heated by the precursor, i.e., photons leaked from the shock and traveling ahead of the shock, after the shock breakout and hydrogen in the dense CSM is ionized. Thus, the dense CSM becomes optically thick and the photosphere is located in the unshocked dense CSM. In this phase, the photospheric velocities are low because the photosphere is in the unshocked CSM. The photospheric velocities gradually increase in our models because the precursor pushes the dense CSM and the unshocked CSM is slightly accelerated. The detailed hydrodynamic structure at these phases can be found in Fig.~6 of \citet{moriya2011iipcsm}.

When the shock reaches the photosphere of the unshocked dense CSM, the photospheric velocity and temperature suddenly increase. After the velocity jump, the photospheric velocity matches the shock velocity for a moment. After the photosphere recedes in the ejecta, the photospheric velocities and temperatures of the models with and without the winds become similar unless the dense CSM has a significant effect on the ejecta dynamics as in the case of $\gtrsim 10^{-2}~\Msunpyr$.

The delayed increase in the photospheric velocities could be a characteristic of the dense CSM around SNe~IIP. However, the photospheric temperatures during the low velocity phase are very high ($\gtrsim 10,000~\mathrm{K}$, bottom panels in Fig.~\ref{fig:photo}) and we expect rather featureless spectra in these epochs. Indeed, SN~2013fs has featureless spectra up to around 10~days when the photospheric velocity could have been low and it is difficult to measure the photospheric velocities in these epochs \citep{yaron2017iipcsm}. In the case of a low-luminosity SN~IIP 2016bkv with a dense CSM, an increase in the photospheric velocity may have been observed because of the relatively small effect of the interaction \citep{hosseinzadeh2018sn2016bkv}.

\section{Discussion and conclusions}\label{sec:discussion}
We have presented our LC models from the RSG progenitors with the CSM in which the acceleration of the wind is taken into account. All the LCs are available at \url{https://goo.gl/o5phYb}. The CSM density at the immediate vicinity of the progenitors can be much higher than the density estimated by assuming a constant wind velocity to the stellar surface. Therefore, the change in the density structure due to the wind acceleration significantly affects the early LCs of SNe~IIP.

The $\beta$-velocity law we adopted is one simple way to present the CSM structure around the progenitors. We adopt the simple formula partly because the wind acceleration mechanism in RSGs is not known. The $\beta$-velocity structure describes the CSM structure of some RSGs like $\zeta$ Aurigae well but it is probably not good in other cases like Antares \citep{ohnaka2017resolvedantares}. The dense CSM as we ascribe the wind acceleration may not even be from the wind acceleration \citep{dessart2017earlyiip}. However, it is worth noting that $\beta$ estimated from the early SN~IIP LCs by applying our LC models \citep{forster2017hits} is consistent with the RSG wind acceleration \citep[e.g.,][]{schroeder1985windbeta}, i.e., $\beta\gtrsim 1$. \citet{grafener2016flashmodel} also found $\beta\simeq 3.5$ for the progenitor of Type~IIb SN 2013cu.

To distinguish these different possible mechanisms to make dense CSM around the SN~IIP progenitors, the spectroscopic observations immediately after the explosions will be helpful. The CSM density estimated based on the spectroscopic observations at 6~hours after the explosion of SN~IIP 2013fs \citep{yaron2017iipcsm} is found to be consistent with our wind acceleration interpretation \citep{moriya2017windacc13fs}. Further similar observations are required to see the diversity in the circumstellar environment of the RSG SN progenitors.

\citet{gonzalez2015iiprise} found that the rise times of SNe~IIP in the optical bands are about $5-10$~days irrespective of the filters (see also \citealt{rubin2016iiprise}). Looking at Figs.~\ref{fig:risepropsdss} and \ref{fig:risepropkepler}, we can find that the rise times become around $5-10$~days in all the optical bands in the models $\Mdot\gtrsim 10^{-3}~\Msunpyr$. This is consistent with the analysis by \citet{forster2017hits} that present 26 rising SN~IIP LCs and find that the mass-loss rates of the immediate SN~IIP progenitors are $10^{-3}-10^{-2}~\Msunpyr$. If the mass-loss rates are higher than $10^{-2}~\Msunpyr$, the rise time increases because of the larger diffusion times in the dense CSM. Therefore, the mass-loss rates higher than $10^{-2}~\Msunpyr$ are not preferred. For the same reason, the CSM radii are not likely to be much larger than $10^{15}~\mathrm{cm}$ in order to keep the rise time in $5-10$~days with $10^{-3}-10^{-2}~\Msunpyr$ (Section~\ref{sec:radii}). The estimated mass-loss rates are much higher than those found in RSGs \citep[e.g.,][]{goldman2017agbrsgwind} and indicate that some mechanisms to enhance mass-loss in the final stages of the RSG progenitors exist.

An important difference caused by the wind acceleration in the mass-loss property shortly before the RSG explosion is the mass-loss period. If the wind velocity is assumed to be constant at 10~\kmps, the mass-loss enhancement needs to continue for decades to reach $10^{15}~\mathrm{cm}$. However, if the wind is accelerated, the mass-loss enhancement needs to be sustained for centuries to reach $10^{15}~\mathrm{cm}$ because the wind is slower at first. The proper estimates for the duration of the mass-loss enhancement would be important information to reveal the mass-loss mechanisms shortly before the explosions \citep[e.g.,][]{quataert2012wavemassloss,shiode2014wavemassloss,fuller2017rsgwavemassloss,moriya2014neutrino}.

The number of the early SN~IIP LC observations will dramatically increase in the coming decade and our LC models will help deriving physical properties of the progenitor systems from such SN~IIP LCs as demonstrated by \citet{forster2017hits}.

\section*{Acknowledgements}
TJM is supported by the Grants-in-Aid for Scientific Research of the Japan Society for the Promotion of Science (16H07413, 17H02864).
FF acknowledges support from Basal Project PFB-03. FF acknowledges support from the Ministry of Economy, Development, and Tourism's Millennium Science Initiative through grant IC120009, awarded to The Millennium Institute of Astrophysics (MAS). FF acknowledges support from Conicyt through the Programme of International Cooperation project DPI20140090.
SCY was supported  by the Korea Astronomy and Space Science Institute under the R\&D
program (Project No. 3348- 20160002) supervised by the Ministry of Science and
ICT, and  by the Monash Center for Astrophysics via
the distinguished visitor program. 
GG is supported by the Deutsche Forschungsgemeinschaft, Grant No. GR 1717/5.
TJM and FF thank the Yukawa Institute for Theoretical Physics at Kyoto University, where this work was initiated during the YITP-T-16-05 on ``Transient Universe in the Big Survey Era: Understanding the Nature of Astrophysical Explosive Phenomena".
Numerical computations were carried out on PC cluster at Center for Computational Astrophysics, National Astronomical Observatory of Japan.






\bibliographystyle{mnras}
\bibliography{ms} 








\bsp	
\label{lastpage}
\end{document}